%% file: cosyr.tex
\documentclass[%
reprint,
amsmath,amssymb,
aps,
prstab,
showkeys
]{revtex4-2}

\usepackage{graphicx}
\usepackage[ruled,vlined]{algorithm2e}
\usepackage{enumitem}
\usepackage[english]{babel}

\begin{document}



\input{Cosyr_Manuscript}

\bibliography{bib/cosyr}

\end{document}

%% file: Cosyr_Manuscript.tex
\title{CoSyR: a novel beam dynamics code for the modeling of synchrotron radiation effects}
\author{C.-K. Huang} \email{huangck@lanl.gov}
\author{F.-Y. Li, H. N. Rakotoarivelo, B. Shen, J. Domine, B. Carlsten, G. Dilts, R.Garimella, T. Kwan, R. Robey} 
\affiliation{Los Alamos National Laboratory, Los Alamos, 87545, NM, US}

\date{\today}
\begin{abstract}
The self-consistent nonlinear dynamics of a relativistic charged particle beam interacting with its complete self-fields is a fundamental problem underpinning many of the accelerator design issues in high brightness beam applications, as well as the development of advanced accelerators. Particularly, synchrotron radiation induced effects in a magnetic dispersive beamline element can lead to collective beam instabilities and emittance growth. A novel beam dynamic code is developed based on a Lagrangian method for the calculation of the particles’ radiation near-fields using wavefront/wavelet meshes via the Green’s function of the Maxwell equations. These fields are then interpolated onto a moving mesh for dynamic update of the beam. This method allows radiation co-propagation and self-consistent interaction with the beam in the simulation at greatly reduced numerical errors. Multiple levels of parallelisms are inherent in this method and implemented in our code CoSyR to enable at-scale simulations of nonlinear beam dynamics on modern computing platforms using MPI, multi-threading, and GPUs. CoSyR has been used to evaluate the transverse and longitudinal coherent radiation effects on the beam and to investigate beam optics designs proposed for mitigation of beam brightness degradation in a magnetic bunch compressor. In this paper, the design of CoSyR, as well as the benchmark with other coherent synchrotron radiation models, are described and discussed.
\end{abstract}

\keywords{Synchrotron radiation \& free-electron lasers, Beam dynamics, Beam code development \& simulation techniques, Electromagnetic field calculation}

\maketitle

\section{Introduction}
\begin{figure}[h]
        \centering
        \includegraphics[width=0.8\columnwidth]{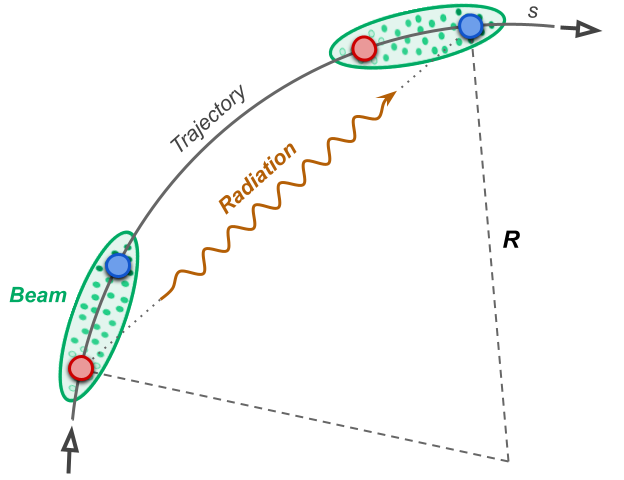}
        \caption{A beam at the current location (geen, top right) in a curved trajectory $s$ of radius $R$ interacts with its radiation. For a particular pair of particles in the beam, e.g., the blue/red particles, the radiation may be emitted at a previous location (lower left) but travels in a shorter straight path to the receiving particle. The catch-up distance is large compared to the radiation wavelength due to the small velocity difference between the relativistic beam and radiation fields in vacuum.}
        \label{fig:CSR}
\end{figure}

The continuing quest to enhance X-ray Free Electron Lasers’ (FELs) performance/functionality and the need for compact advanced accelerators demand techniques to manipulate electron beams with the highest brightness (i.e., the beam density in 6D phase space) possible. However, nonlinear beam dynamic problems often arise in the generation and control of such beams. In particular, an electron beam emits an electromagnetic wave in the form of synchrotron radiation when accelerated, e.g., by external fields in a magnetic section — a common building block for many critical beam line components including beam compressor, beam cooler, emittance exchanger, undulator, etc.. It is well known \cite{Schwinger1949, MURPHY1997} that a beam of length $\sigma_s$ and $N_p$ electrons emits synchrotron radiation both coherently and incoherently in a curved trajectory $s$ of radius $R$. The coherent power has dependence on the number of electrons and beam length as $P_{coh} \sim N_{p}^2 \alpha^{-4/3}$, and the incoherent power depends on the energy and number of the electrons as $P_{incoh} \sim N_{p} \gamma^4$, where $\gamma$ is the Lorentz factor of the beam and $\alpha=\sigma_s/R$ is its angular width. The synchrotron radiation emitted in the forward direction catches up with the beam, after a long distance $l \sim \alpha^{1/3} R \gg \sigma_s$, and shines on it (Fig. \ref{fig:CSR}). The interaction with the coherent radiation can interfere with the macroscopic motion of different parts of the beam, resulting in effects such as beam self-steering \cite{Allison1998} and  collective instabilities. In particular, the longitudinal Coherent Synchrotron Radiation (CSR) effect can lead to amplification of the initial beam current modulation (microbunching), primarily via longitudinal energy modulation and the magnetic dispersion $R_{56}$ \cite{Stupakov2002, Huang2003}. Furthermore, the transverse component of the coherent radiation can directly change the beam emittance \cite{Jing2013}, $\epsilon$, an important beam parameter measuring its area in the $(x, v_x)$ or $(y, v_y)$ transverse phase space. On the other hand, the incoherent effect generates random shot noises and phase space diffusion \cite{Garcia:FEL2017-TUP024}. These can lead to detrimental emittance growth and also affect the modulation/seeding of beams for high harmonics generation in FEL. Therefore, the nonlinear interaction between the beam and its synchrotron radiation self-fields can cause significant degradation of quality for a high brightness beam. It has been recognized \cite{Barletta2010} that these nonlinear effects represent major physical obstacles for FEL performance enhancement and advanced experimental techniques for beam phase space manipulation. The above scalings  indicate that coherent and incoherent synchrotron radiation effects are most pronounced at high peak current (large $N_{p}$ and small $\alpha$) and high energy (large $\gamma$), respectively, which are both present for a beam needed to drive a hard X-ray FEL or dense energetic beams produced in advanced accelerators (e.g., $N_{p} \sim 10^{9}$, $\alpha \sim 10^{-5}$, $\gamma \sim 10^{3}$). Hence, it is vital to study the consequence of both the collective and stochastic nonlinear interactions for applications involving high-brightness beams. In particular, it is important to understand how strongly the longitudinal and transverse fields would coherently or statistically couple to the instability and emittance growth, and their relative importance. This knowledge will enable the development of mitigation strategies and the optimization of a particular design.

Despite constant increase in computing power enabling simulations with one macroparticle per electron of a realistic beam, existing simulation models are (1) not capable of capturing both coherent and incoherent synchrotron radiation effects at the same time and (2) also lacking the self-consistency and/or accuracy required by the problem. This is due to the disparate spatial scales, high frequency of the radiation and the history dependent nature of the emission. Furthermore, a peculiar connection between the relativistic formulation and geometrical configuration causing cancellation of terms requires judicious choice of numerical methods and accuracy \cite{Cai2017}.

The coherent synchrotron radiation models in standard beamline design tools only take the longitudinal effect into account in its most basic form \cite{Bassi2006}. Most such models employ the Liénard-Wiechert potential, i.e., the retarded Green’s function of the Maxwell equations of a moving particle. The “steady-state” assumption \cite{Saldin1997a} is often made, in which the beam travels into a circular trajectory for a sufficient time so that the electromagnetic fields around the beam are all emitted from this circular trajectory. However, the time difference in emission is ignored when considering the beam dynamics, i.e., radiation propagation and interaction between any particle pairs are instantaneous. The concept of synchrotron radiation wakefields can be readily applied with this simplification.  It is further assumed that all beam particles follow the same reference trajectory for the emission process, so the steady-state radiation wakefields for a reference particle can be applied to every particle in order to avoid brute-force calculation at high cost $\sim O(N_{p}^{2}N_{t})$, where $N_{t} >10^{3}$ is the number of time steps required for typical simulations. For an evolving beam, this approach is not self-consistent as the history dependence of the emission from each individual particle is ignored and all interactions are treated as instantaneous despite that the evaluation of the wakefields still requires the retarded time. Hence, it is only suitable for describing the stage of instability growth when the evolution of the driving field is mostly determined by the beam density profile. These models are usually implemented in a 1D mean-field approximation (assuming a pencil beam with smooth profile) and only for the longitudinal dynamics from the coherent fields while excluding both incoherent and transverse effects. A multi-dimensional steady-state model has also been developed where the retarded time of the emission on the reference trajectory is solved for each mesh points in a mesh co-moving with the beam \cite{Ryne:IPAC2018-THPAK044}. A history search is necessary for the given reference trajectory, which can be done by solving a quadratic equation for the retarded time in the case of a circular trajectory. Nonetheless, this equation is quite nonlinear and a highly accurate solution is needed for the Green’s function due to the multi-scale nature of the radiation fields.

Particle-mesh models via the discretization of the full-wave Maxwell equations, e.g., the Finite Difference Time Domain (FDTD) method, are popular for electromagnetic modeling of the beam dynamics and also employed \cite{Novokhatski2011,Fawley2010} for this problem. They are, in principle, self-consistent for the coherent effects, but in practice their accuracy is severely limited by the numerical dispersion error from propagating the high-frequency radiation on a mesh over a catch-up distance much longer than the radiation wavelength (e.g., Fig. \ref{fig:CSR}). Additionally, the spatial/temporal errors for the Lorentz force in the particle pusher due to the staggered mesh commonly used (e.g., in the Yee scheme) is a concern. To ensure an accurate numerical dispersion for modeling the coherent synchrotron radiation of a short beam, it necessitates the high cost of a second-order FDTD solver which scales with  $\alpha^{-4/3}$, even when a moving simulation window is used. The paraxial model \cite{Gillingham2007a} can avoid the high-frequency fields by evolving the envelope of the radiation fields, but the validity of this approximation is restricted to a narrow emission cone. These models have yet to demonstrate reliable results on the coherent effects for practical beam parameters, due partly to the unfavorable cost scaling for accurate simulations and partly to model limitations. In addition, the granularity of the particle distribution and the multi-scale nature of the radiation are lost when particle currents are interpolated to the mesh with a cell size much larger than the corresponding length scales.

In light of this need, we are developing a unique code, CoSyR \cite{github_zenodo}, as a versatile and accurate simulation tool to tackle the fundamental problem of the nonlinear dynamics of a particle beam from its self-fields, particularly the radiation fields, which underpins many accelerator design issues in high-brightness beam applications, as well as those arising in the development of advanced accelerators. Compared with the standard beamline design tools that only take the longitudinal coherent effect into account in its most basic form \cite{Bassi2006}, CoSyR includes both the longitudinal field and the transverse field, which are essential to correctly simulate the interplay between the two, such as the beam "self-steering" effect \cite{Allison1998}, and the emittance growth from the transverse fields radiations that are important for high-current-density beams. Furthermore, the “steady-state” assumption \cite{Saldin1997a} for the Liénard-Wiechert potential employed in most existing beam dynamics models, which require all particles to follow the same reference trajectory during the emission process, is relaxed, so detailed beam dynamics and instability can be studied. In CoSyR, we use a flexible method to separate the close-by (non-paraxial) emission and those from farther locations (mostly paraxial), allowing both emissions from independent trajectories of the particles while simplifying the calculation for the former events. CoSyR also couples a Green's function based solver with the particle-mesh method to avoid brute-force calculation at high cost $\sim O(N_{p}^{2}N_{t})$ . As will be detailed in section \ref{sec:code}, for each particle, their fields (or potential) are calculated on the radiation wavefronts that intersect with a common moving mesh. These wavefronts are emitted at a specified interval as the particle travels along the trajectory. Similar to the scheme by Shintake \cite{Shintake2003}, $N_{w}$ discrete points (denoted “wavelets”) on the sections of wavefronts overlapping with the common moving mesh are chosen to naturally adapted to the emission. The wavelets are further divided into two groups -- dynamic and subcyled, depending on the retarded time and the characteristic timescale of the beam evolution. By using a moving mesh of $N_{m} \sim O(N_{w})$ mesh points to accumulate the fields emitted from the particles and to update the beam, the cost of the CoSyR field solver is $\sim O(N_{p}N_{m}N_{t})$, while its particle pusher has a similar cost $\sim O(N_{p}N_{t})$ as in the Particle-In-Cell method. CoSyR is also implemented using the MPI + Kokkos \cite{CarterEdwards2014} programming model designed and optimized for the era of exascale-computing to allow at-scale beam dynamic simulations, e.g., on heterogeneous CPU-GPU platforms.

In this paper, we will focus on the modeling of coherent radiation fields, and defer the discussion of incoherent radiation fields to a future publication.

\section{General features and geometry self-similarity of the radiation field}

\begin{figure}[h]
        \centering
        \includegraphics[width=1.0\columnwidth]{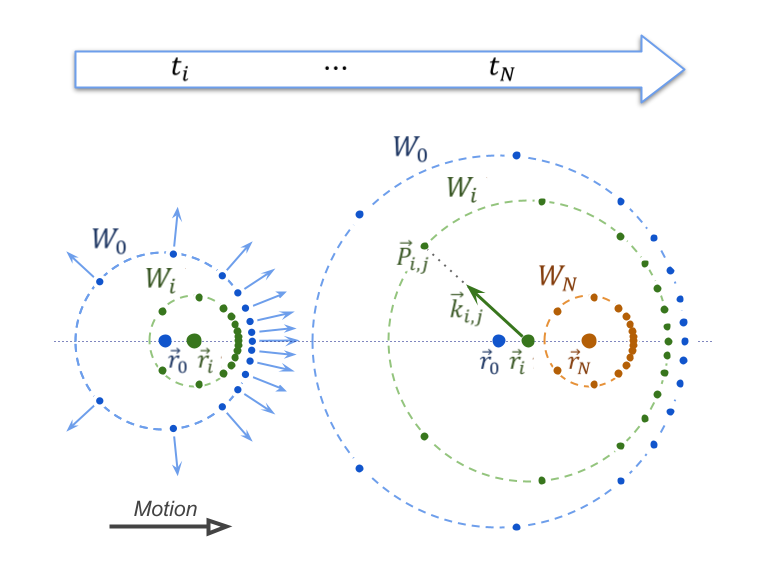}
        \caption{The concept of wavefront/wavelet generation and tracing for a moving electron initially located at $\vec{r}_{0}$ and subsequently moves to $\vec{r}_{i}$ and $\vec{r}_{N}$ at time step $t_i$ and $t_N$. $W_0$, $W_i$ and $W_N$ label the expanding wavefronts emitted at those locations. The solid dots on the wavefronts represent the wavelets.}
        \label{wavefront-tracing}
\end{figure}

Without loss of generality, we will use the cyclotron motion as an example in the following sections. The unit used is defined by the base quantities $(L, T, V, M)$ for length, time, velocity and mass in the cgs system. A particularly relevant unit system for cyclotron motion is $(L, T, V, M) = (R, R/c, c, m_e)$  where $R=\gamma m_e v c /q B_0$ is the cyclotron radius for an electron of energy $\gamma m_e c^2$ under a constant magnetic field $B_0$ and $c$ is the light speed in vacuum. In this system, the cyclotron radius is unity and the unit of the electric/magnetic field is $m_ec^2/eR = B_0/\gamma \beta$.

CoSyR exploits a physical property of radiated fields from an electron (i.e., the radiation wavefronts)  to simplify the coherent and incoherent field calculation, which in turn enables a self-consistent dynamics calculation. The design of CoSyR is further motivated by the general features of the radiation field. These features are best illustrated by the expanding radiation wavefronts/wavelets emitted from a moving electron, as introduced in the work of Shintake \cite{Shintake2003} and shown in Fig.~\ref{wavefront-tracing}. Specifically, at time step $t_i$, the electron at the position $\vec{r}_i$ would emit a wavefront $W_i$, containing a set of wavelets denoted by $\vec{P}_{i,j}$, where $j$ represents the wavelet index for the direction of emission. At subsequent time steps $t_N$ ($t_{N}>t_{i}$), the wavefront (and the associated wavelets) would simply propagate outward as 
\begin{equation}
        \vec{P}_{i,j}(N\Delta t)=\vec{r}_i+c(N-i)\Delta t \vec{k}_{i,j},
\end{equation}
where $\Delta t$ is the time step size and $\vec{k}_{i,j}$ is the unit propagation vector measured relative to the origin of emission, i.e., $\vec{r}_i$. Notice that, the above notation essentially tags the outermost wavefront as $W_0$, thereby it avoids shifting all emitted wavefronts in memory at each time step as opposed to the original formalism where the innermost (i.e., newly emitted) wavefront is tagged as $W_0$.

\begin{figure*}[ht!]
        \centering
        \includegraphics[width=0.8\columnwidth]{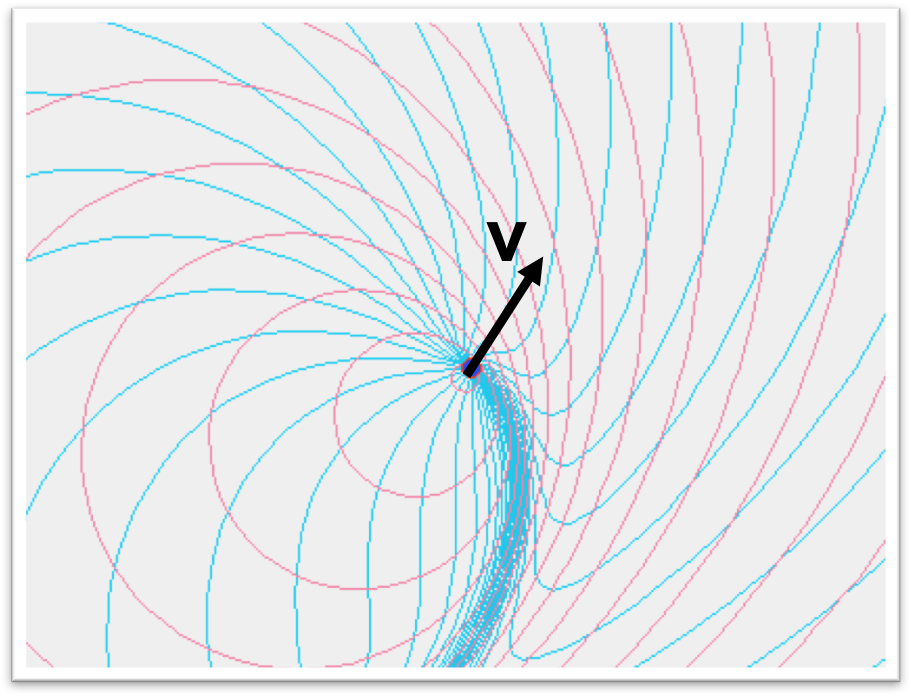}
        \includegraphics[width=1.0\columnwidth]{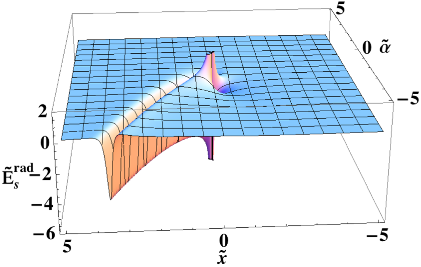}        
        \caption{(Left) The wavefronts (red circles) and the trajectories of a fixed set of wavelets chosen according to Shintake’s scheme (blue curves) from an emitting electron of instantaneous velocity $\vec{v}$. The clustered wavelets indicate the location of the incoherent field traced by the wavefronts and wavelets. (Right) Multi-scale self-similar structure of the longitudinal radiation near field $E_s^{rad}$ in 2D for an electron in uniform circular motion. The field structure consists of a “trough” (orange surface) and a smooth region (blue surface). The self-similar scaling is given by $\tilde{E}_s^{rad} \equiv E_s^{rad} R^2/(e\gamma^4)$ in the Frenet-Serret coordinates scaled by the Lorentz factor $(\tilde{x} \equiv x \gamma^2/R, \tilde{y} \equiv y \gamma^2/R, \tilde{\alpha} \equiv s\gamma^3/R)$. The emitting particle is at the origin and its motion is in the $x-\alpha$ plane. The “trough” feature in 2D becomes a “needle” in 3D with opening $\Delta{\alpha} \sim \gamma^{-3}$ and $\Delta y \sim \gamma^{-2}/R$ \cite{Huang2013PAC, Garcia2017PHD}. Similar features and scaling exist for the transverse fields, and scalar/vector potentials.}
        \label{fig:mesh_field}
\end{figure*}

The wavelets $\vec{P}_{i,j}$ represent the locations where the Green’s function of electromagnetic fields or potentials can be easily evaluated, e.g., through the Liénard-Wiechert radiation field equations \eqref{eq:field-only-kernel-E} and \eqref{eq:field-only-kernel-B}, without the need to solve for the retarded time. It is proposed in Ref.~\cite{Shintake2003} that one can equally divide the $2\pi$ angle of the initial wavefront in 2D (or $4\pi$ solid angle in 3D) to define the corresponding emission directions in the instantaneous rest frame of the electron. Due to the relativistic beaming effect, the resulting wavelets will then cluster around the direction of the electron velocity after the Lorentz transform to the laboratory frame,  i.e.,
\begin{equation}
        \label{eq:k_transform}
        k_x=\frac{\cos\theta'+\beta}{1+\beta\cos\theta'}, k_y=\frac{\sin\theta'}{\gamma(1+\beta\cos\theta')},
\end{equation}
where $\theta'$ is the uniformly distributed emission angle in the electron frame measured from the direction of $\vec{\beta}$ and $\vec{v}=\vec{\beta}c$ the electron velocity in the laboratory frame. 
By tracing the emissions at successive steps, one obtains a set of near-field wavefronts/wavelets that display strong spatial-scale variations, i.e., the wavelets are bunched in a narrow region and sparsely distributed in other area (Fig. \ref{fig:mesh_field}).

A similar scale separation in radiation fields was also discovered recently through a geometric self-similarity analysis \cite{huang2013two}. In general, the near-field radiation consists of a relatively weak but large-scale feature and a strong but narrow trough-like feature (needle-like in 3D) \cite{Huang2013PAC, Garcia2017PHD} shown in Fig. \ref{fig:mesh_field}. The former is responsible for the coherent effects, while the latter is mostly responsible for the incoherent effects. We note that the latter feature can extend far beyond the emitting electron in one particular direction \cite{huang2013two}; therefore, its collision-like interaction with other particles is \textit{long-range but unilateral}. In the local Frenet-Serret coordinates $(x, y, s)$, where $s$ is the coordinate along the particle trajectory and $x, y$ are the coordinates in the plane perpendicular to the tangential direction ($x$ is in the bending plane), both the radiation field strength and its spatial profile scale with the particle’s Lorentz factor and the curvature of the trajectory $1/R$. In Fig. \ref{fig:mesh_field}, $\alpha=s/R$ denotes the angular offset in the tangential (longitudinal) direction, and $\chi=x/R$ is the radial offset from the reference trajectory; both the coordinates and the field amplitude are properly scaled with $\gamma$, as a result of a self-similarity feature ~\cite{huang2013two}. Most prominently, the scale separation with wavefronts/wavelets found in Shintake’s scheme allows us to use them to capture these features without resorting to a prohibitively expensive global fine mesh. In this work, we use the wavefronts/wavelets to compute the coherent field by remapping the fields of wavelets to a moving mesh; the collision-like interaction through the spiky feature will be included in the future.

\section{Beam dynamics code: CoSyR}\label{sec:code}

\subsection{Overview}

\begin{figure}[h]
        \centering
        \includegraphics[width=1.0\columnwidth]{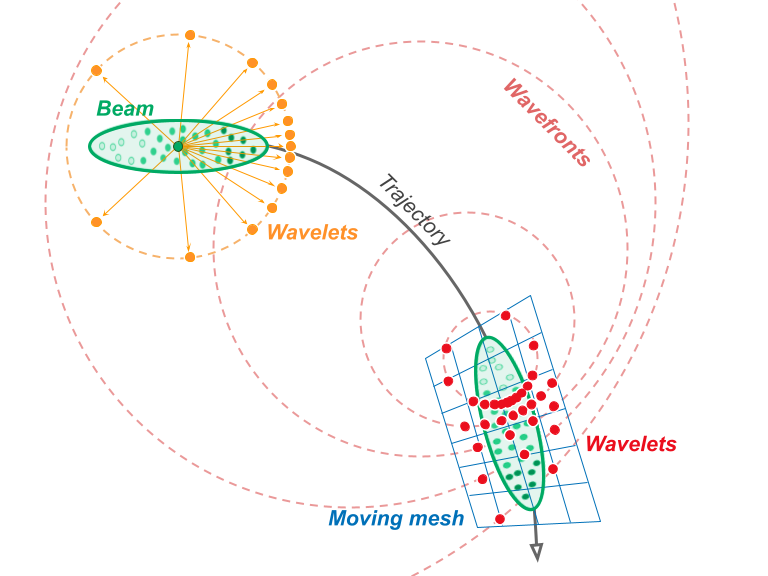}
        \caption{A schematic for the key concepts in CoSyR. A particle beam moving along the trajectory in the black curve at two instances of time are shown in green ovals and dots. The wavefronts emitted by a particular beam electron are shown as red or orange dots. The orange dots represent the scheme proposed by Shintake for the choices of the wavelets. In CoSyR, only those wavelets (red dots) overlapping with the moving mesh are generated at the present time.}
        \label{fig:concept}
\end{figure}

CoSyR is a high-performance C++ code with exascale simulation capability/design in mind. The key concepts in CoSyR, including wavefronts, wavelets and the moving mesh are illustrated in Fig.~\ref{fig:concept}. The code consists of three major components: a field/wavelet computation kernel for each electron, a mesh remapping module to aggregate the fields from the wavelets onto the moving mesh, and a particle pusher that uses the aggregated fields to advance the electrons. Unlike other particle-mesh codes with a local PDE-based field solver where communication only occurs between neighboring MPI ranks, CoSyR's field solver is based on the retarded Green's function and thus is nonlocal both in time and space. This feature allows both decoupling of the time/spatial scales in coherent and incoherent effects, and improved accuracy for the solution to the beam self-fields.

In CoSyR,  a technique similar to the overset grid used \cite[Chapter~9]{Meakin1999} for fluid flow simulations around moving bodies is used to overlay individual electron's wavefronts/wavelets with the moving mesh representing the simulation domain around the beam (Fig. ~\ref{fig:concept}). The moving mesh follows a reference particle in the beam which has the design energy and trajectory. The wavelets on the wavefronts are used to store values of the Green’s functions (the electric and magnetic fields and/or wakefield potentials) near the present location of each electron, while the moving mesh stores the accumulated values for all electrons in the beam. Since the wavefronts/wavelets from an electron never intersect each other due to causality, they can be simply stored in a regular data structure. The generation of the wavelets is designed to avoid the disparate resolution requirements for resolving the fields, and their management (e.g., retiring wavefronts from the computation) is straightforward, as will be discussed in the next section. The field values at the wavelets (i.e., source points) around each mesh point (i.e., target point) are used to reconstruct the values on the moving mesh, employing a highly accurate local regression technique \cite{Fan1996} which allows adaptive sampling in the source wavelets. Given a set of points with field data, the local regression estimator can simultaneously compute both a best fit function and its derivatives on the set of target points using an arbitrarily high-order polynomial or a B-spline basis. The wavelet-to-mesh interpolation is implemented using an optimized hybrid-parallel (MPI+X) remapping library, Portage \cite{Portage}. Portage allows adaptive, kernel-density estimation or high-order local regression for transferring field data between the wavelets and the moving mesh points.

The moving mesh currently used in CoSyR is uniform, but a structured or unstructured mesh can also be implemented if needed, thanks to the flexibility in Portage. In turn, the collocated electric and magnetic fields (or the longitudinal/transverse wakefields) on the moving mesh are used to push particles for their coherent interaction with the large-scale radiation fields, as in a Particle-In-Cell code. This is done without the error usually associated with staggered meshes, which can be important especially for relativistic particles that experience near canceling fields in the Lorentz force. The particle pusher is similar to those used in existing high-performance kinetic plasma simulation codes, such as the VPIC code \cite{Bowers2008}. It is also implemented through a performance portable library, Cabana \cite{Cabana}, from the Co-Design Center for Particle Applications (CoPA) within the Exascale Computing Project supported by DOE.

While CoSyR is built upon the work of Shintake to evolve the positions of the wavelets and to calculate the radiation fields, several modifications/improvements have been implemented. In the following, we briefly outline these changes and present more details in the next few sections.

\begin{enumerate}
        \item Shintake's original scheme for wavelet generation is modified to restrict the wavelets to the sections of wavefront intercepted by the moving mesh at its current location, leading to a significant reduction of the wavelets needed; 
        \item The scale separation in wavelet distribution can lead to sparse coverage of the large-scale coherent field region. An improvement is made to distribute the wavelets on a particular wavefront uniformly in the lab frame instead of the electron frame; 
        \item We have identified the missing term in Shintake's method for field calculation \cite{li2019validation}. Including such term in Shintake's method is found to be equivalent to the Liénard-Wiechert retarded solution, but with considerable computation (see Appendix). Therefore,  we directly use the latter for the calculation of the fields and/or potentials; 
        \item Furthermore, the wavelets are divided into two groups depending on the retarded time taken for their propagation and the fastest timescale of the dynamics of interest. When the retarded time is smaller than the characteristic timescale (which typically needs to be resolved by the simulation time step), these wavelets and associated fields/potentials are pre-calculated by ignoring the difference in the velocities of the emitting particles. They are used in the simulation together with the dynamic wavelets generated for the emissions at larger retarded time, as will be discussed in the following section. This approach amounts to treating the dynamics faster than the characteristic one using the “steady-state” approximation, which is justified by the short propagation distance of the radiation field. This further reduces the computation cost, while fully self-consistent fields are retained for the beam dynamics resolved by the time step.

\end{enumerate}

\begin{algorithm}[h]
        \SetKwBlock{DoParallelMPI}{do in parallel with MPI}{end}       
        \SetKwBlock{DoParallelOMP}{do in parallel threads}{end}
        \DoParallelMPI{
                initialization\;
                \While{time $<$ total simulation time}{
                        push reference particle\;
                        update moving mesh\;
                        \DoParallelOMP{
                            push other particles\;
                                \eIf{wavelet emission}{
                                        update emission information\;
                                }{
                                        skip emission\;
                                }
                                shift subcycle wavelets\;
                                update dynamic wavefronts\;
                                \eIf{wavefront intersection with moving mesh}{
                                        compute fields for the wavelets on the dynamic wavefronts\;
                                }{
                                        skip field calculation\;
                                }   
                        }    
                        \DoParallelOMP{                
                                interpolation to the moving mesh\;
                            }    
                                field/potential summation for all non-reference particles\;
                            gradient estimate (if needed)\;
                }
        }
        \caption{Algorithm of CoSyR}
\label{algorithm}
\end{algorithm}

The flow chart of the CoSyR code is presented in Algorithm \ref{algorithm}. Multiple levels of parallelisms are exploited in CoSyR, as particles are completely independent of each other, as well as the wavelets emitted by the particles and the mesh points. Since the self-fields of each electron is strictly independent, the outer loop over electrons is "embarrassingly parallel"; in our current implementation it is primarily parallelized over particles through MPI processes at the top level as well as CPU/GPU threads at the second level. The key steps, namely wavelet/field calculation, field remapping from the wavelets to the moving mesh, and finally the particle push using the fields at the mesh, are described separately in more detail in the following sections.

\subsection{Generation of Wavelets}

While the original wavelet generation proposed by Shintake was shown to be useful for animating the near fields, several modifications have to be made in order to apply it to CoSyR. Firstly,  the relativistic-beaming effect (Eq.~\eqref{eq:k_transform}) may cause the wavelets to aggregate outside the moving mesh, leaving just a few wavelets to cover this mesh. It can also lead to strongly nonuniform wavelet distributions in the laboratory frame, posing challenges in the subsequent remapping step.

\begin{figure}[h]
        \centering
        \includegraphics[width=1.0\columnwidth]{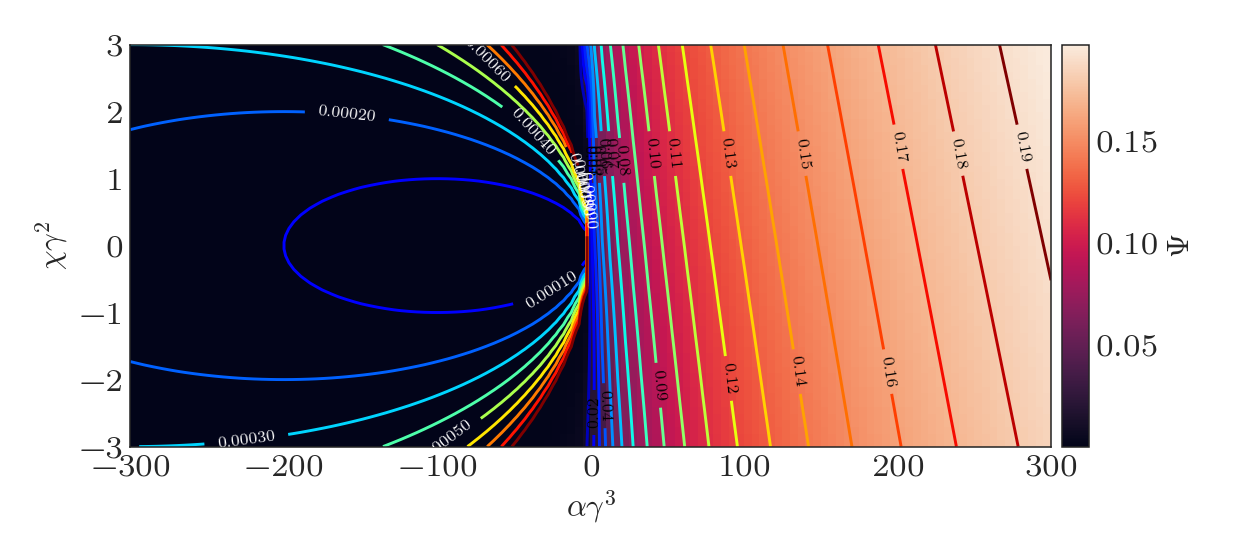}
        \caption{Map of the retarded angle/time $\Psi$ in the coordinates of the comoving mesh ($\alpha, \chi$), properly scaled with electron energy $\gamma$ for the case of $\gamma=100$ and $R=1$m. The contour lines show constant $\Psi$ at equal separations of $\Delta\Psi=1\times10^{-2}$ in the right region and $\Delta\Psi=1\times 10^{-4}$ on the left. 
        }
        \label{psi-contour}
\end{figure}

Secondly, as another consequence of the highly relativistic electron motion, the timescale of the emissions that finally contribute to the moving mesh also varies drastically with spatial locations. This can be seen from the retarded angle ($\Psi$)  on the moving mesh calculated for a reference electron, as presented in Fig.~\ref{psi-contour}. T        he reference electron is always located at the center of this mesh, $(\alpha,\chi)=(0,0)$ (see the schematic Fig. 1 of Ref.~\cite{huang2013two}). The retarded angle, $\Psi$, is related to the retarded time (i.e., the time needed for the emission generated at the retarded position to propagate to the current location) by $t_{E}=\Psi /\beta\sim\Psi$. According to Fig.~\ref{psi-contour} which is typical for an emitting relativistic electron, the retardation is normally large in front of the electron (i.e., $\alpha>0$), but becomes extremely small behind the electron ($\alpha<0$). Hence, for a fixed emission time interval $\Delta t_E$,  fewer wavefronts fill in the left region of the moving mesh than in the right region. It should be noted that for the region of small $\Psi$ values, the field may not be negligible (except for around the axis $\chi=0$) and it contributes substantially to the coherent radiation for a beam. Such a field is also responsible for the fastest beam evolution due to the near-instantaneous interaction between an electron residing in this region and the emitting electron. Furthermore, the emission angle for these wavelets relative to the instantaneous electron velocity can be as large as $2\pi$ (see, for example, the dark blue complete contours in the left region of Fig.~\ref{psi-contour}, therefore the paraxial approximation is not applicable in this region.

In order to overcome the above issues, we implement in CoSyR two measures to adjust the wavelet generation. To address the wavelet distribution outside the moving mesh due to the relativistic beaming effect, we restrict the wavelet positions to the moving mesh and adjust the emission directions directly in the laboratory frame. To see how this works, we append on the $\Psi$ map (Fig.~\ref{psi-contour}) contour lines, which represent constant $\Psi$ at equal separations of $\Delta\Psi\sim \Delta t_E$. These contours should be viewed as the wavefronts generated in CoSyR when $\Delta \Psi=\Delta t$ as each contour has the same $\Psi$ value. Using the original scheme for wavelet emission, the wavelets would be distributed highly nonuniformly along the contours and are not constrained within the moving mesh which only occupies a small region in the whole space. In CoSyR, we drop the transform of emission direction from the electron to the laboratory frame, but focus on the segments of the wavefronts that intersect the moving mesh (i.e., the contours shown in the $\Psi$ map in Fig.~\ref{psi-contour}). For simplicity, we let the electron emit uniformly in angle within each segment. While this approach allows for much more uniform wavelet distribution within the moving mesh, it necessitates a dynamical adjustment of the wavelet generation based on the current location of the moving mesh. The wavelet angular distribution only needs to be dynamically adjusted for the wavefronts within the moving mesh. In particular, when $t_E>\Psi_{max}$ ($\Psi_{max}$ is the maximum value of $\Psi$ on the mesh), the wavefronts will outrun the mesh and not contribute to the beam dynamics, therefore no wavelet needs to be generated or adjusted. To determine whether a particular wavefront outruns the moving mesh, one can simply check for the intersection of the wavefront with the mesh boundary. The maximum number of wavefronts for dynamic wavelet generation is given by

\begin{equation}
        N_{max}= \frac{\Psi_{max}}{\beta\Delta t}.
\end{equation}

The above method adjusts wavelet distribution along the contours, i.e., it mostly improves the distribution in the radial $\chi$ direction. The wavefront distribution along the longitudinal $\alpha$ axis can be controlled by the emission interval $\Delta t_E$ in the simulation. Since this time interval is a pseudo time used to discretize the field and does not alter the actual field distribution, $\Delta t_E$ can be chosen such that wavelets in the right region are manageable. However, it is not possible to adjust the inclination of the wavefront intersections (or the contours) as it is inherent to the wavefront emission scheme. Compared to the approach where trajectory history search or analytic/numerical solver is used to obtain the retarded angle for the specified mesh points, the above method for wavelet generation can be viewed as a way to quickly identify the range of the retarded angles for the mesh points near the wavefronts, while field remapping is used instead of attempting to find the exact solution to the retarded angle. We also note that the mesh points may reside in some particle’s incoherent field zone, but the coherent field can still be properly accounted for even if so, when the wavelet generation is adjusted for such a zone.

To address the extremely small emission time $t_E$ in the left region ($\alpha<0$), we can "subcycle" the emission (i.e., retarded) time to generate more wavelets. The extremely small retarded time means that the position of the emitting electron is very close by, and the electron would have little variation in position and velocity during such a small time interval. Therefore, we can provide predefined subcycle wavelets instead of the dynamically generated ones through an approximation for the region where $\Psi$ value is small. As discussed earlier, the Green’s functions can be evaluated on these wavelets assuming the emitting particle has the same velocity as the reference particle. This assumption is justified as the subcycle wavelet emission direction can be significantly different from the velocity directions of the emitting and the reference particle, therefore one can ignore the difference of the latters. These wavelets, including their positions and field values, are calculated separately following Ref. ~\cite{huang2013two} for the reference electron and supplied into CoSyR as an input. They are currently distributed uniformly and are loaded into the simulation at a grid density similar to the moving mesh grid to ease remapping onto the latter. These subcycle wavelets will be reused for other electrons but shifted accordingly (see Algorithm 1). A crucial element is to find out the timescale corresponding to these subcycle wavelets. It could be taken as the minimum between the following two quantities,
\begin{equation}
        \label{t_sub}
        t_{sub}=\rm min\left\{\Delta t, \frac{R\Delta\alpha}{4\beta}+\frac{R\Delta\chi^2}{2\beta\Delta\alpha}\right\},
\end{equation}
where the latter is essentially the timescale for the wavefronts to just reach the left corners of the moving mesh. For a large mesh scale (or beam scale), the latter could be larger than the simulation step $\Delta t$, and hence the subcycle wavelets only partially occupy the left region (close to the mesh center). In practice, the subcycle timescale could be taken to be multiple times of $\Delta t$ by filling a larger region on the left with precalculated wavelets regardless of the above condition in Eq.~\eqref{t_sub}. This crude approximation could result in partial overlap between the subcycle wavelets and those dynamically generated, and its effects on the convergence of the beam field have been found to be small.

\subsection{Kernel calculation}

Once the wavelets, including both the dynamic and subcycle ones, are generated, we compute the values of the Green’s function on these wavelets. We have currently developed two formalisms for the kernel, one based on the Liénard-Wiechert fields directly and the other based on a mixed formalism with the transverse Lorentz force and the potential of the longitudinal wakefield. Both are derived from the Liénard-Wiechert solution which is validated in the Appendix, but only the so-called “acceleration term” is included in the fields at present. In CoSyR, a local Cartesian coordinate at the present location of the reference particle is used for the kernel calculation and particle update instead of using a specific curvilinear coordinate defined on the reference trajectory.

In the first formalism, denoted “field-only”, the following terms in the Liénard-Wiechert fields associated with the acceleration are used,

\begin{align}
\vec{E}^{rad} &= - \frac{e\hat{n} \times [(\hat{n}-\vec{\beta}') \times \dot{\vec{\beta}}'] }{c r'(1-\hat{n}\cdot\vec{\beta}')^3}, \label{eq:field-only-kernel-E}\\
\vec{B}^{rad} &= \vec{n} \times \vec{E}^{rad}.  
\label{eq:field-only-kernel-B}
\end{align}

Here, $\vec{\beta}'$ and $\dot{\vec{\beta}}'$ are stored along the trajectory and $\vec{n}$ is chosen according to our wavelet generation scheme. The retarded distance $r'$ that usually requires a trajectory search or the solution of a quartic equation is simply $r' = c(N-i) \Delta t$ as described before.

Eqs. ~\eqref{eq:field-only-kernel-E} and ~\eqref{eq:field-only-kernel-B} are general and can be used in  3D simulations. Below,  our formalisms are described using the 2D geometry in the bending plane to illustrate the properties of the kernels. In 2D, $\vec{E}^{rad}$ is projected along the longitudinal and transverse directions of the reference trajectory as $E_{s}^{rad}$ and $E_{x}^{rad}$ respectively, while  $\vec{B}^{rad}$ reduces to a component perpendicular to the bending plane, i.e., $B^{rad} \hat{y}$. When $\vec{\beta}' \cdot \dot{\vec{\beta}}' = 0$ and $\hat{\beta}' \times \dot{\vec{\beta}}' = -\beta'^2c/R \hat{y}$, these components can be simplified as follows,

\begin{align}
E^{rad}_{s} &=  \frac{e \beta'^{2} (1 - \hat{n} \cdot \vec{\beta}' - \gamma'^{-2})}{R^2 \Psi (1-\hat{n}\cdot\vec{\beta}')^3}  (\hat{n} \cdot \hat{x}) \label{eq:field-only-kernel-Es}\\
E^{rad}_{x} &=  \frac{e \beta'^{2} (1 - \hat{n} \cdot \vec{\beta}' - \gamma'^{-2})}{R^2 \Psi (1-\hat{n}\cdot\vec{\beta}')^3}  (\hat{n} \cdot \hat{s}) \label{eq:field-only-kernel-Ex} \\
B^{rad} &= \frac{e \beta'^{2} (1 - \hat{n} \cdot \vec{\beta}' - \gamma'^{-2})}{R^2 \Psi (1-\hat{n}\cdot\vec{\beta}')^3}
\label{eq:field-only-kernel-By}
\end{align}
where we use $(\hat{n} - \vec{\beta}') \times \dot{\vec{\beta}}' = -(\hat{\beta}' \times \dot{\vec{\beta}}') (1 - \hat{n} \cdot \vec{\beta}' - \gamma'^{-2}) / \beta'$,  $ \hat{n} \times [ (\hat{n} - \vec{\beta}') \times \dot{\vec{\beta}}'] = -\hat{n} \times (\hat{\beta}' \times \dot{\vec{\beta}}') (1 - \hat{n} \cdot \vec{\beta}' - \gamma'^{-2}) / \beta'$ and $ \hat{n} \times \hat{n} \times [ (\hat{n} - \vec{\beta}') \times \dot{\vec{\beta}}'] =  (\hat{\beta}' \times \dot{\vec{\beta}}') (1 - \hat{n} \cdot \vec{\beta}' - \gamma'^{-2}) / \beta'$ .

The “field-only” formalism is straightforward but there are two numerical issues that limit its usage: (1) the fields are highly spiky due to both the $(1-\hat{n}\cdot\vec{\beta}')^3$ factor in the denominator and the $(1 - \hat{n} \cdot \vec{\beta}' - \gamma'^{-2})$ factor in the numerator, e.g., as shown in Fig. \ref{fig:mesh_field} for the longitudinal field. In particular, the width of the spiky fields, which can be estimated from the conditions $\hat{n} \parallel \vec{\beta}'$ and $ \hat{n} \cdot \vec{\beta}' = \beta'^2$, has an unfavorable spatial scaling with the particle energy, hence requiring a fine moving mesh resolution in order for the accumulated beam fields to converge;  (2) there is significant cancellation in the transverse Lorentz force $F_{x} = -e(E_{x}^{rad} - \beta B^{rad})$, as can be seen from Eqs. \eqref{eq:field-only-kernel-Ex} and \eqref{eq:field-only-kernel-By}, that requires the corresponding components of the electric and magnetic fields to be evaluated at sufficient accuracy. However, those fields are always collocated at the wavelets or the mesh points hence discrepancy in remapping accuracy for different field components is not a concern. Despite the high mesh resolution needed, we demonstrate that this formalism still works for relatively low-energy beams and can serve as a baseline implementation for the general simulation setup with fewer approximations.

In the second formalism, these two numerical issues are alleviated through a pseudo-potential for the longitudinal acceleration field and an approximation of the transverse Lorentz force. We call this the “mixed-kernel” formalism. Following earlier works, the pseudo-potential can be defined as $(\phi - \beta A_{s})$ (see, e.g., \cite{huang2013two}), and $E_{s}^{rad} = - \partial_{s} (\phi - \beta A_{s})$ is calculated after remapping and accumulation on the moving mesh. In the transverse direction, the Lorentz force is $F_{x, j} = -e(E_{x}^{rad} - \beta_{j} B^{rad})$ and generally particle dependent with $\vec{\beta}_{j}$ being the individual particle velocity and $\vec{\beta}_{0}$ being the reference particle velocity. However, when $| \Delta\beta_{j} B^{rad}| = | (\beta_{j} - \beta_{0}) B^{rad} | \ll |(E_{x}^{rad} - \beta_{0} B^{rad})|$, which may be justified for a high brightness beam with very small energy spread, the transverse Lorentz force (per electron) can be reduced to an effective transverse wakefield  $W_{\perp} = (E_{x}^{rad} - \beta_{0} B^{rad})$. This approach takes advantage of the large cancellation within the Lorentz force and the accumulation of the kernel can converge at a mesh resolution less stringent than that for the "field-only" kernel.

\subsection{Remapping}

\subsubsection{Interpolating fields from the wavelets to the mesh}
\begin{figure}[hb]
\centering
\includegraphics[width=0.8\columnwidth]{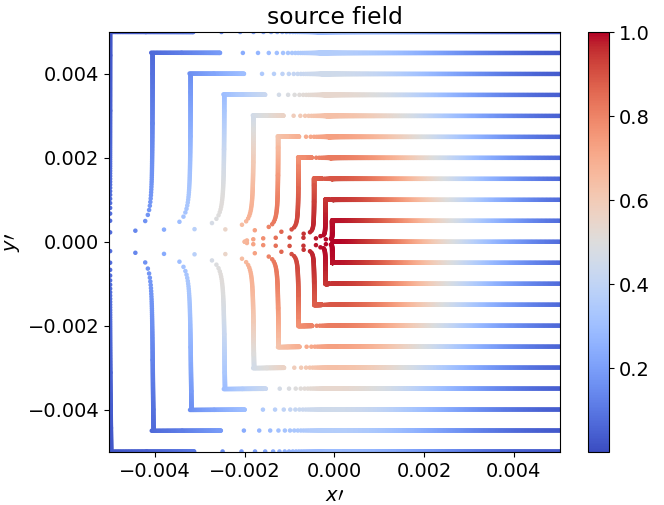}
\includegraphics[width=0.8\columnwidth]{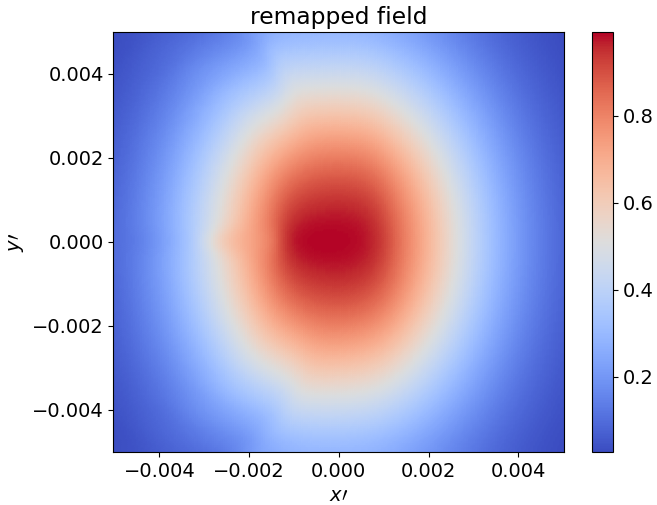}
\caption{Example of remaping of a two-dimensional Gaussian field from discrete wavelets (top) to a cartesian grid (bottom) using a local regression method. It shows that even if the wavelets are sparsely distributed on the domain, the field is correctly reconstructed on the regular cartesian grid.}
\label{fig:remap-example}
\end{figure}

At this point, the goal is to interpolate and accumulate the field values computed on the wavelet points emitted by all particles onto the points of the moving mesh. The electric and magnetic fields from the mesh will be used to push the particles for their coherent interaction. In our case, the field values are approximated on the mesh using local regression estimation. Basically, the idea is to infer the function values at each mesh point by finding its best approximation using arbitrary high-order polynomials. Since the estimation is local, an adaptive sampling of the wavelet points can be used to reduce the computational cost while keeping the same order of accuracy.
A simple example of a local regression-based remap is given in Figure~\ref{fig:remap-example}.


Let us denote $f:\Omega \to \mathbb{R}$ the field to be interpolated from wavelets to mesh points. It may represent either the longitudinal or the transverse radiation field. For the sake of clarity, we explain the method for the 1D case, but the logic is the same in 2D. Here, wavelets and mesh points are called \textit{source} and \textit{target} points respectively.
The idea is then to compute the approximated solution $f^h(x_{\mathsf{t}})$ at a target point $x_{\mathsf{t}}$ from the exact values $f(x_{\mathsf{s},i})$ at set of source neighbors~$\{x_{\mathsf{s},i}\}_{i=1}^n$ using a weighted moving least square fit. In this case, the local solution at any point $x$ in the vicinity of $x_{\mathsf{t}}$ can be approximated with:

\begin{align}
f^h(x) & = \sum_{j=0}^m b_j(x_{\mathsf{s},t}) c_j \\
& = \mathbf{b}(x) \cdot \mathbf{c}
\label{eq:principle-least-squares}
\end{align}

with:
\begin{itemize}[noitemsep]
 \item  $f^h$ the approximation of $f$ in the vicinity of $x_{\mathsf{t}}$.
 \item $\mathbf{b}(x) = (1,x,\frac{x^2}{2}, \cdots, \frac{x^m}{m!}) = [b_0(x), \cdots, b_m(x)]$ \newline a vector of \textit{basis} functions.
 \item $\mathbf{c}$ a vector of \textit{constants} coefficients.
\end{itemize}
\noindent

To compute \eqref{eq:principle-least-squares}, we have to find the coefficients $\mathbf{c}$ of the polynomial fit using a weighted least squares method. For that, we aim to minimize a sum of weighted squared residuals, with respect to the coefficients $\mathbf{c}$:

\begin{align}
\min_{\mathbf{c}} \mathsf{S} & = \min_{\mathbf{c}} \sum_{i=1}^n w_{\mathsf{t}}(x_{\mathsf{s},i} - x_{\mathsf{t}}) (f^h(x_{\mathsf{s},i}) - f(x_{\mathsf{s}, i}))^2
\label{eq:residual-mls}
\end{align}

\begin{figure}[h]
\centering
\includegraphics[width=0.5\columnwidth]{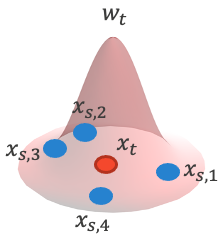}
\caption{Example of a weight function defined at a mesh point $x_\mathsf{t}$. Its value smoothly decreases as the distance between the target point $x_\mathsf{t}$ and its source neighbor $x_{\mathsf{s}, i}$ increases.}
\label{fig:remap-weight}
\end{figure}

\noindent
Here the weight functions $w_{\mathsf{t}} : \Omega \to \mathbb{R}$ allow us to control the contribution of any source point $x_{\mathsf{s},i}$ according to its distance from $x_{\mathsf{t}}$ (and to discard those that are not in its local vicinity). In fact, $w_{\mathsf{t}}$ is maximal when both points coincide, and smoothly decreases as their distance increases, as shown in Figure~\ref{fig:remap-weight}.
Here we discard any point $x_{\mathsf{s},i}$ lying outside the local vicinity of $x_{\mathsf{t}}$, which is delimited by the \textsf{support} of $w_\mathsf{t}$~\footnote{the support is equivalent to the \textit{smoothing lengths} in \textsf{Smooth Particle Hydrodynamics} methods.}, given as an input.

%
%

The minimization problem in \eqref{eq:residual-mls} can be addressed by setting each partial derivative of $\mathsf{S}$ to zero with respect to each coefficient $c_k$ of $\mathbf{c}$.
%
%
%
%
%
To find those coefficients, we then have to solve the $m$ normal equations in \eqref{eq:normal-equations}.

\begin{align}
\forall k, \partial_{c_k} \mathsf{S}_{(\mathbf{c})}  = 0  & \Leftrightarrow 2 \mathbf{b}_k \mathbf{W}(x_{\mathsf{t}}) (\mathbf{B} \mathbf{c} - \mathbf{f}) = 0\\
& \Leftrightarrow \mathbf{b}_k \mathbf{W}(x_{\mathsf{t}}) \mathbf{B} \mathbf{c}  = \mathbf{b}_k \mathbf{W}(x_{\mathsf{t}}) \mathbf{f}
\label{eq:normal-equations}
\end{align}

\noindent
with $\mathbf{b}_k$ the vector of the $k^\mathsf{th}$ basis function values at each~$x_{\mathsf{s},i}$, $\mathbf{W}$ a diagonal matrix of weights functions values, $\mathbf{B}$ the matrix of all basis functions values at each~$x_{\mathsf{s},i}$, and $\mathbf{f}$ the exact function values at each~$x_{\mathsf{s},i}$:
\begin{align}
\forall k, \mathbf{b}_k^T & = [b_k(x_{\mathsf{s},1}), \cdots, b_k(x_{\mathsf{s},n})]\\
\mathbf{W}(x) & =
\begin{bmatrix}
w_\mathsf{t}(x_{\mathsf{s},1} - x) & \cdots & 0\\
\vdots & \ddots & \vdots\\
0 & \cdots & w_\mathsf{t}(x_{\mathsf{s},n} - x)\\
\end{bmatrix}\\
\mathbf{B} & = [\mathbf{b}_0, \cdots, \mathbf{b}_m]\\
\mathbf{f}^T & = [f(x_{\mathsf{s},1}), \cdots, f(x_{\mathsf{s},n})]
\end{align}


\noindent
In fact, \eqref{eq:normal-equations} can be rewritten as a single matrix equation:
\begin{align}
\nabla \mathsf{S}_{(\mathbf{c})} = 0 & \Leftrightarrow \mathbf{B}^T \mathbf{W}(x_{\mathsf{t}}) \mathbf{B} \mathbf{c} = \mathbf{B}^T \mathbf{W}(x_{\mathsf{t}}) \mathbf{f}\\
& \Leftrightarrow \mathbf{c} = (\mathbf{B}^T\mathbf{W}(x_{\mathsf{t}})\mathbf{B})^{-1} \mathbf{B}^T\mathbf{W}(x_{\mathsf{t}})\mathbf{f}
\label{eq:least-squares}
\end{align}

At this point, we have a way to compute the constant coefficients $\mathbf{c}$ involved in the approximation of $f^h$ in the vicinity of~$x_\mathsf{t}$. But in reality, we aim to locally reconstruct the values of the function using its Taylor series expansion in the vicinity of~$x_\mathsf{t}$ as shown in Eq. \eqref{eq:taylor-expansion}.

\begin{align}
f^h(x) & = f^h(x_{\mathsf{t}}) + \sum_{k=1}^m \frac{\mathrm{d}^k}{\mathrm{d}x^k}f^h(x_{\mathsf{t}}) \frac{x - x_{\mathsf{t}}}{k!}\\
& =\mathbf{b}(x - x_{\mathsf{t}}) \cdot \beta(x_{\mathsf{t}})
\label{eq:taylor-expansion}
\end{align}

with:
\begin{itemize}[noitemsep]
\item $\mathbf{b}(x) = (1, x, \frac{x^2}{2}, \cdots, \frac{x^m}{m!})$
\item $\beta(x) = [f^h(x), \frac{\mathrm{d}}{\mathrm{d}x}f^h(x), \cdots, \frac{\mathrm{d}^m} {\mathrm{d}x^m}f^h(x)]$
\end{itemize}

\noindent
The coefficients $\beta(x)$ corresponding to the values of the function and its derivatives at $x$ can be obtained using the same least squares approximation we used to compute $\mathbf{c}$, as shown in \eqref{eq:reconstruct-beta}:

\begin{equation}
\nabla \mathsf{S}_{(\beta(x))} = 0 \Leftrightarrow \beta(x) =\underbrace{(\mathbf{B}^T \mathbf{W}(x) \mathbf{B})^{-1} \mathbf{B}^T \mathbf{W}(x)}_{\mathbf{M}(x)} \mathbf{f}
\label{eq:reconstruct-beta}
\end{equation}

\noindent
Once $\beta(x_{\mathsf{t}})$ is computed, the reconstructed value of $f$ at $x_\mathsf{t}$ is obtained by simply taking its first component:
\begin{equation}
f^h(x_\mathsf{t}) = \mathbf{b}(0) \cdot \beta(x_{\mathsf{t}}) = \beta_0(x_{\mathsf{t}})
\label{eq:reconstructed-function}
\end{equation}

In practice, we do not need to compute all the coefficients of the matrix $\mathbf{M}(x)$ in~\eqref{eq:reconstruct-beta} to compute~\eqref{eq:reconstructed-function}.
The remapping is performed on the wavelets of each particle on the same \textsc{mpi} rank. Afterwards, the remapped fields from all particles are summed and broadcasted to the other \textsc{mpi} ranks.

\subsubsection{Estimating the derivatives of the pseudo-potential}

For the mixed formalism, the field is obtained by taking the derivative of the longitudinal pseudo-potential at mesh points. As such, we need to estimate the gradient of the remapped pseudo-potential at mesh points. For a set of wavelets associated to a given particle, the derivatives of the function being remapped could be approximated along with its value using equation~\eqref{eq:reconstruct-beta}. However, it would be computationally expensive because we would estimate the gradients for each particle before accumulating them from all particles, possibly from different \textsc{mpi} ranks. Instead, we reconstruct the gradients of the mesh field using a least squares fit.

Let us denote $f:\Omega \to \mathbb{R}$ the field on the mesh. The idea is to approximate the gradient $\nabla^h f$ at a mesh point $\mathrm{x}_i$, given the field values at a set of neighbors $\{\mathrm{x}_j\}_{j=1}^n$. From the Taylor series expansion of $f$ at each $\mathrm{x}_j$ in the vicinity of $\mathrm{x}_i$, we obtain a set of $n$ normal equations:
\begin{align}
\forall j, f(\mathrm{x}_j) & = f(\mathrm{x}_i) + (\mathrm{x}_j - \mathrm{x}_i)^T \nabla^h f(\mathrm{x}_i) \\
\forall j, \underbrace{f(\mathrm{x}_j) - f(\mathrm{x}_i)}_{\mathbf{F}_j}  & =  \underbrace{(\mathrm{x}_j - \mathrm{x}_i)^T}_{\mathbf{A}_j}  \nabla^h f(\mathrm{x}_i)
\end{align}

\noindent
which can be rewritten as a single matrix equation:
\begin{align}
\mathbf{A} \nabla^h f(\mathrm{x}_i) & = \mathbf{F}\\
\mathbf{A}^T \mathbf{A} \nabla^h f(\mathrm{x}_i) & = \mathbf{A}^T \mathbf{F} \label{eq:symmetrize}\\
\nabla^h f(\mathrm{x}_i) & = (\mathbf{A}^T \mathbf{A})^{-1} \mathbf{A}^T \mathbf{F}
\label{eq:reconstruct-gradient}
\end{align}

\noindent
with $\mathbf{A}$ the matrix of distances between $\mathrm{x}_i$ and each of its neighbors $\mathrm{x}_j$, and $\mathbf{F}$ the vector of differences between field values at $\mathrm{x}_i$ and at each $\mathrm{x}_j$:
\begin{equation}
\mathbf{A} =
\begin{bmatrix}
x_1 - x_i & y_1 - y_i\\
\vdots & \vdots \\
x_n - x_i & y_n - y_j
\end{bmatrix} ,
\mathbf{F} =
\begin{bmatrix}
f(\mathrm{x}_1) - f(\mathrm{x}_i)\\
\vdots\\
f(\mathrm{x}_n) - f(\mathrm{x}_n)
\end{bmatrix}
\end{equation}

Since $\mathbf{A}$ is not a squared matrix, we have to symmetrize it before inverting it. In our context, if $f$ represents the remapped pseudo-potential, then the field is obtained by taking $\partial_x^h f$. Its smoothness depends on the number of neighbors involved in the least squares fit.

\subsection{Particle pusher}

The particle pusher in CoSyR, including the interpolation from the moving mesh to the particle, is of the standard Boris type \cite{birdsall2018plasma}. In the case of "field-only" formalism, both longitudinal, transverse electric fields and out-of-plane magnetic fields are stored at the same set of mesh points and interpolated identically to the particle's location using a linear weight function as in a PIC code. With this approach, error in the Lorentz force due to staggered grids in conventional PIC codes is avoided. As mentioned earlier, Cartesian coordinate is used for the pusher instead of a curvilinear one, hence there are no inertial forces associated with the equation of motion to be discretized. The particle coordinate is global, and their projected coordinates in the moving mesh are used for the field interpolation. The resulting fields are projected back to the global coordinate for particle update. These two extra projections represent only a small portion of the overall cost in the pusher. They are done with machine accuracy at each step, compared to the accuracy of a discretized inertial force that depends on the time step. In practice, we find this approach produces a sufficiently accurate trajectory for the motion in the external field.

For the mixed formalism, the interpolation procedure is similar, with the effective field replacing the electric field in the transverse direction.

\section{Benchmark and beam dynamics simulations with coherent radiation}

\subsection{Benchmark with the 2D steady-state model}

\begin{figure*}[ht!]
        \centering
        \includegraphics[width=0.68\columnwidth]{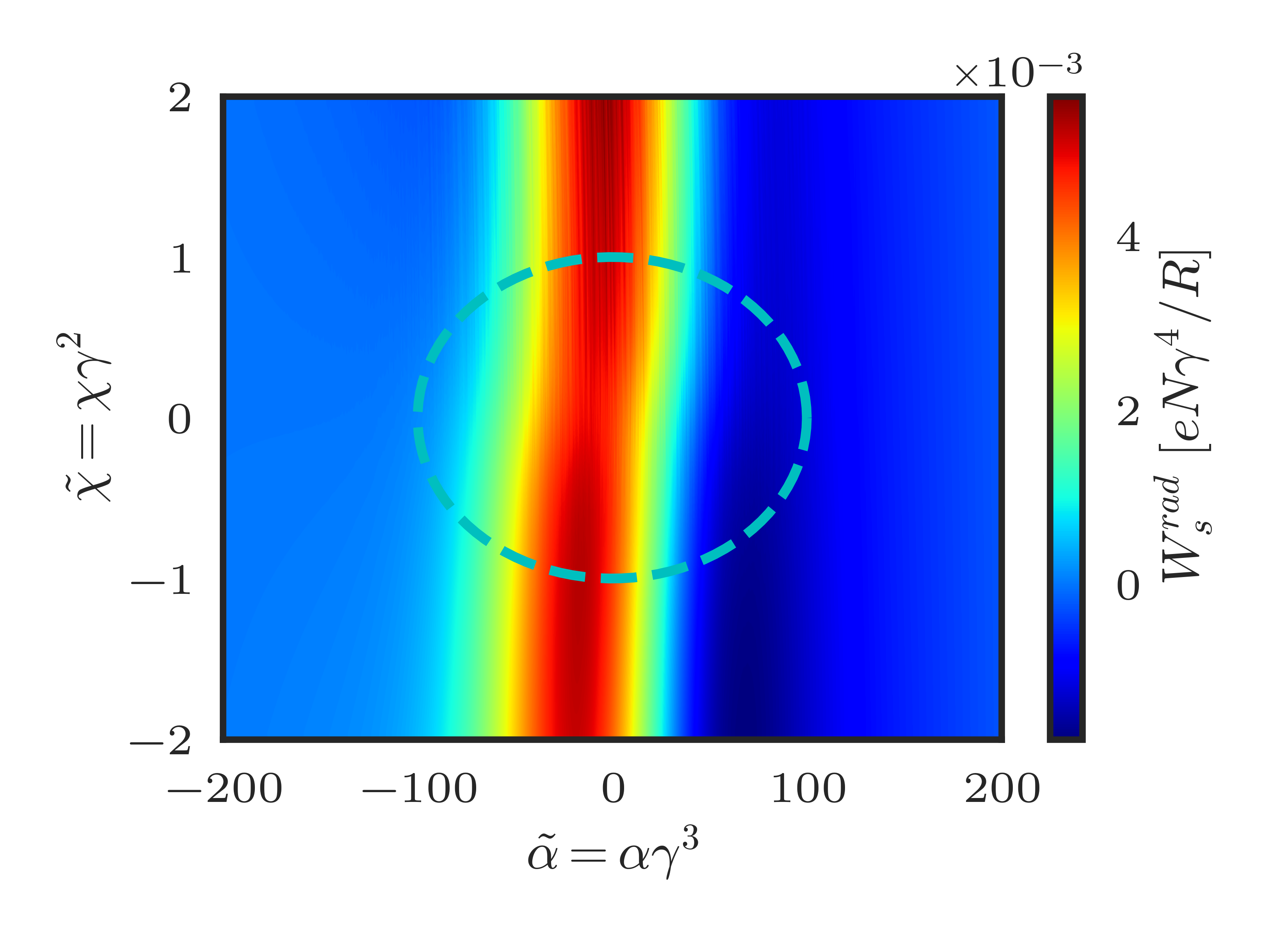}        
         \includegraphics[width=0.68\columnwidth]{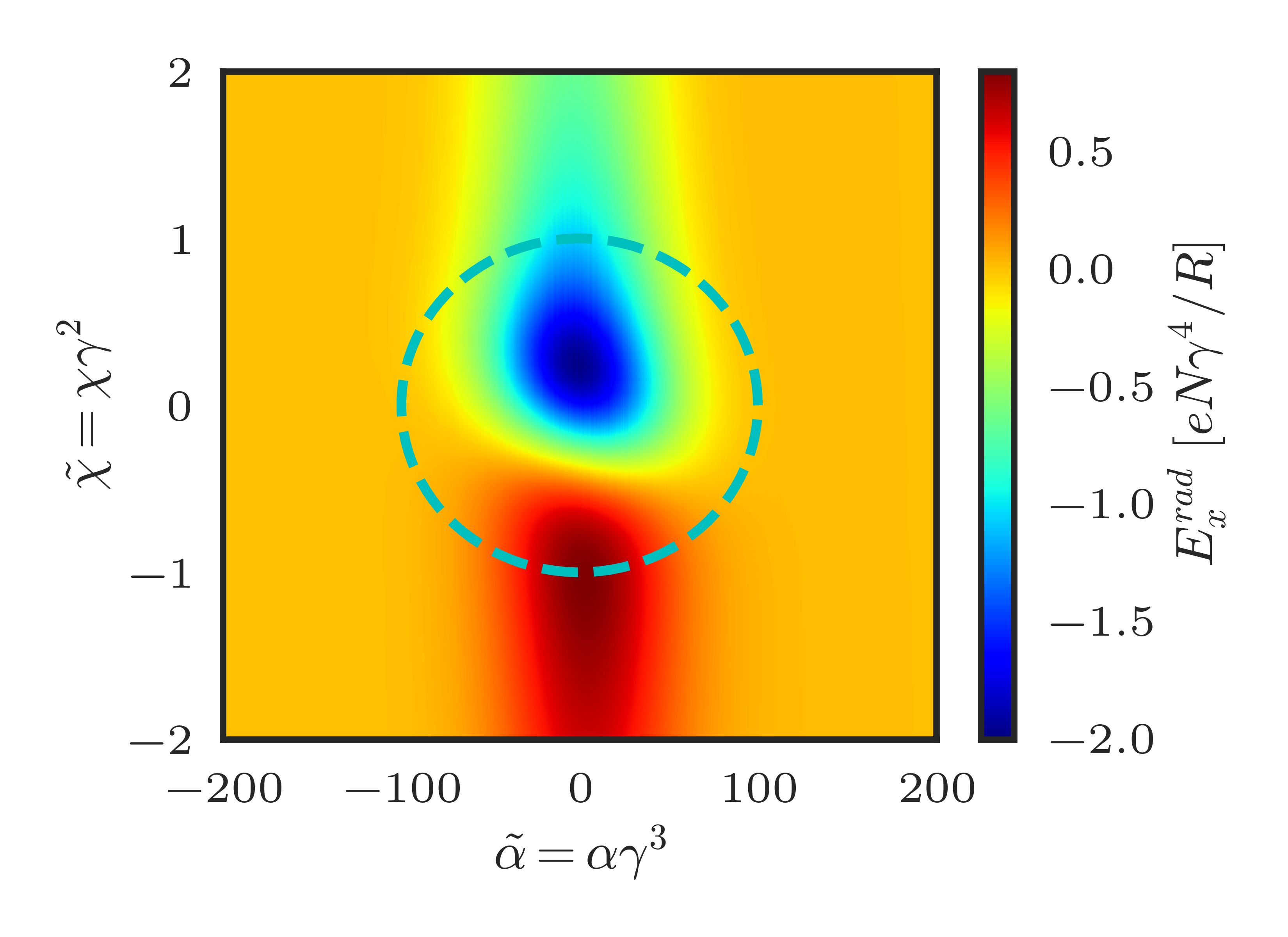}      
         \includegraphics[width=0.68\columnwidth]{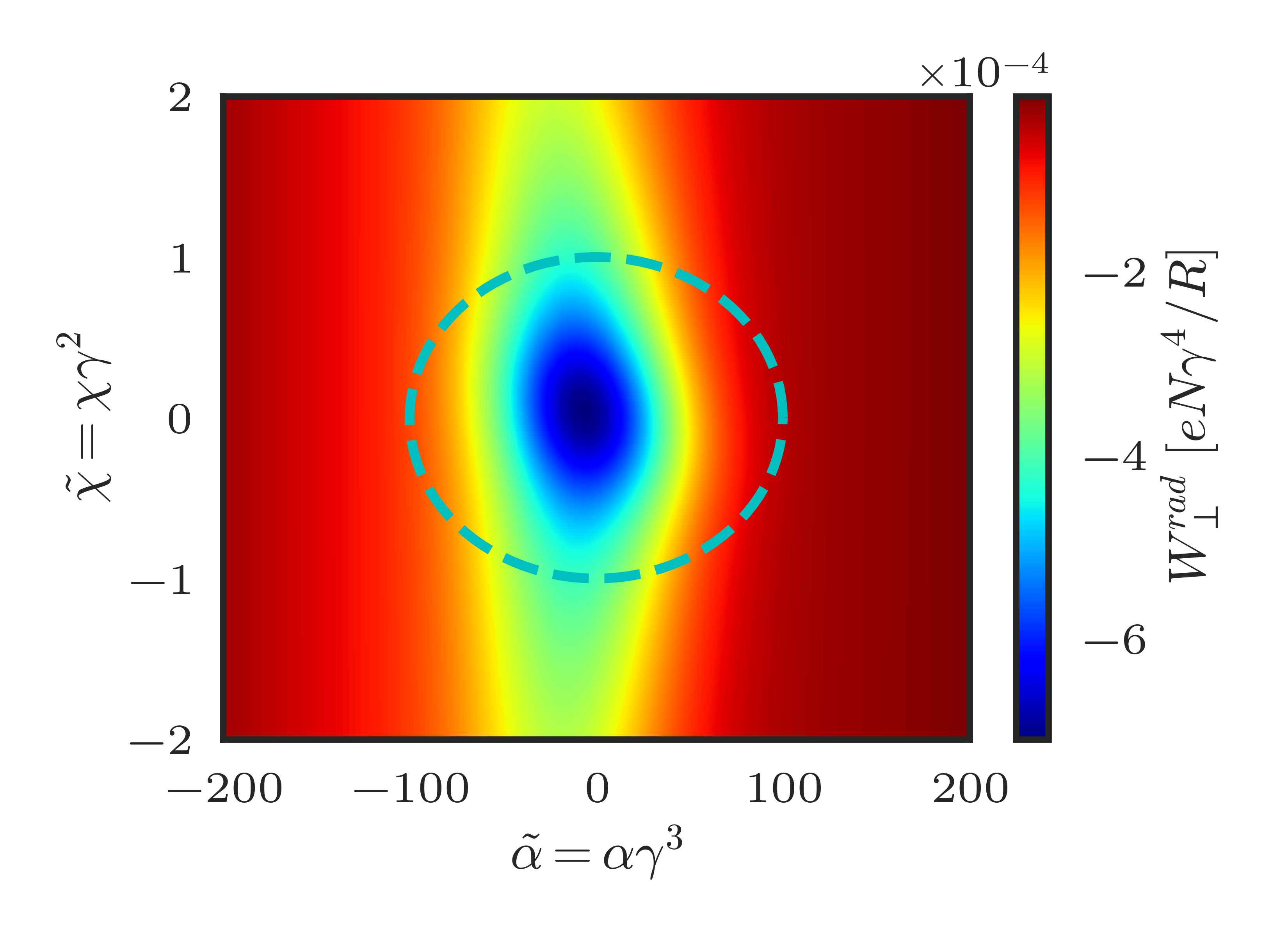}  
        \caption{(From left to right) The steady-state longitudinal (along the direction of the beam motion, denoted by $s$) electric field, transverse electric field and the effective transverse wakefield from the coherent radiation (acceleration field) for an electron beam of $0.01$ nC charge and $200\mu$m spot size at a bending angle of $0.3$ rad in a $R=1$m magnetic dipole section. Only subcycle wavelets are used. The initial beam has a round bi-Gaussian shape (with the green dashed circle of $3\sigma$ radius denoting its boundary), a Lorentz factor of $\gamma=100$ and no energy spread.  In the simulation, $\sim 1.02\times 10^7$ computation particles are used for the beam. Bi-linear smoothing is used for the contour plots. The beam moves to the right in these plots.}
        \label{fig:benchmark-g100}
\end{figure*}

\begin{figure*}[ht!]
        \centering
        \includegraphics[width=0.68\columnwidth]{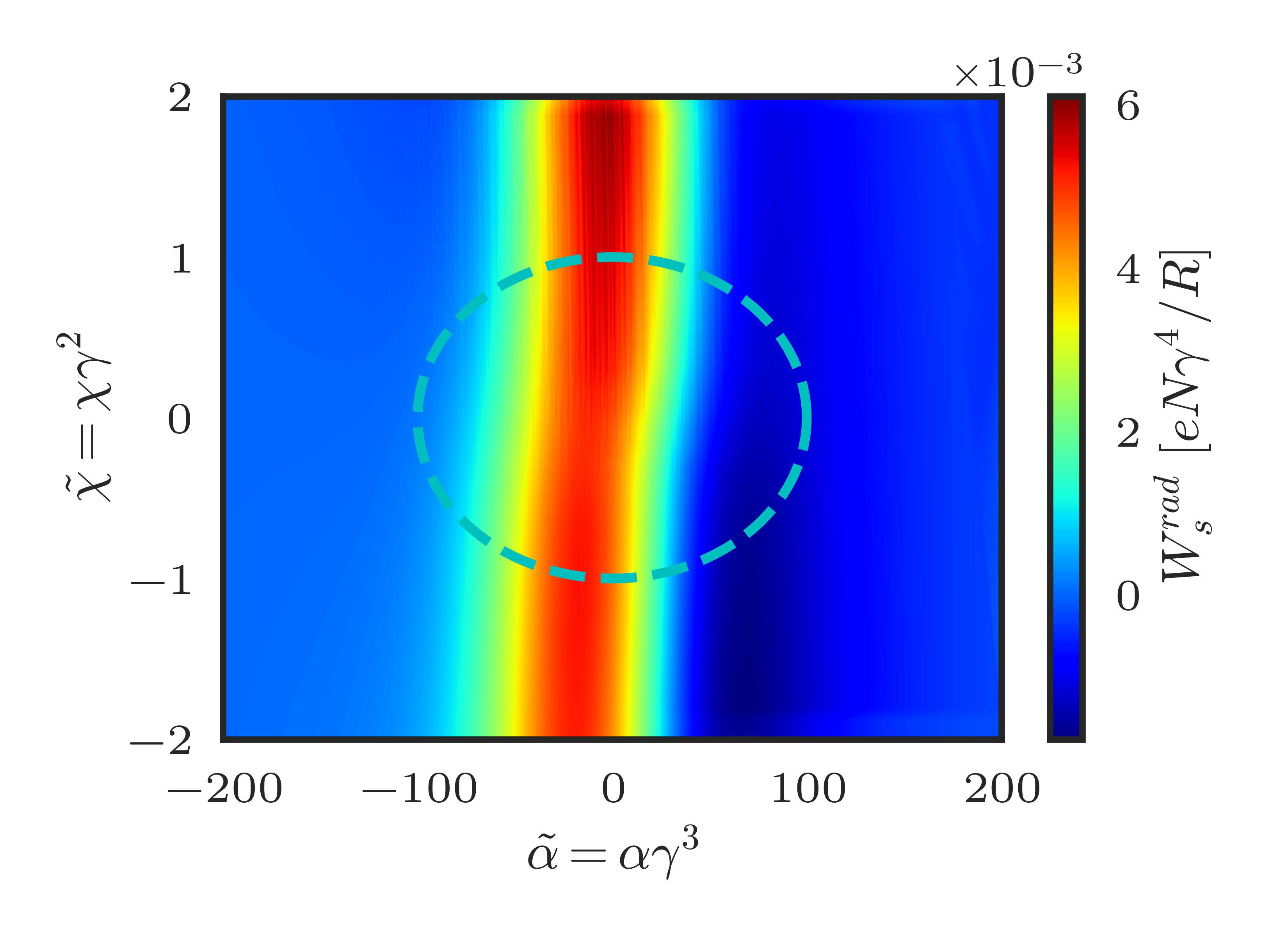}        
        \includegraphics[width=0.68\columnwidth]{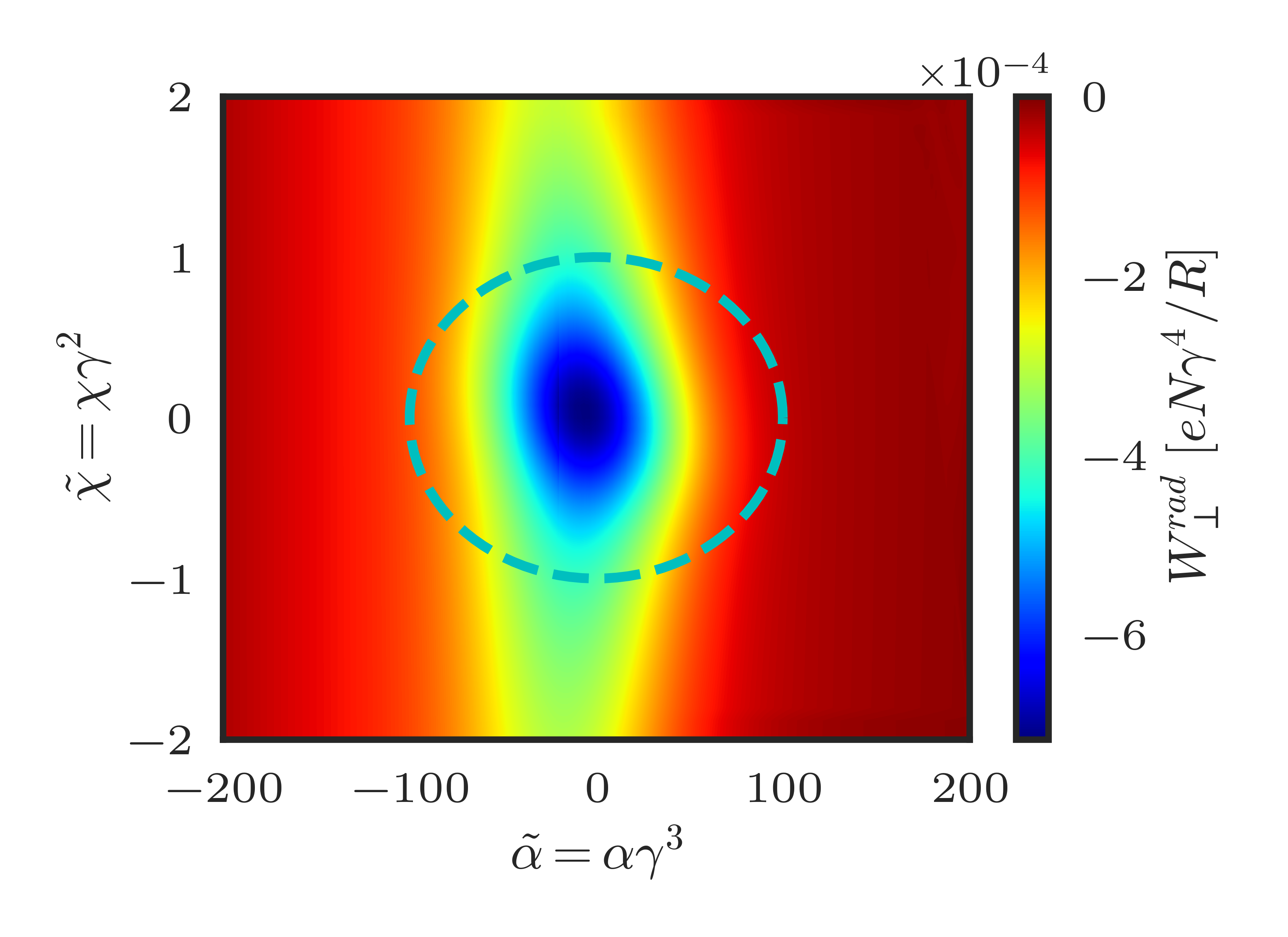}  
        \caption{The non-steady-state longitudinal (left) and transverse wakefields (right) from the coherent radiation (acceleration field) for the same setup as in Fig. \ref{fig:benchmark-g100}, but using both dynamic and subcycle wavelets. There is a noticeable difference with the steady-state longitudinal field in Fig. \ref{fig:benchmark-g100}, see also the comparison in Fig. \ref{fig:LW-CSR-benchmark}. }
        \label{fig:benchmark-g100-dyn}
\end{figure*}

\begin{figure*}[ht!]
        \centering
        \includegraphics[width=0.9\columnwidth]{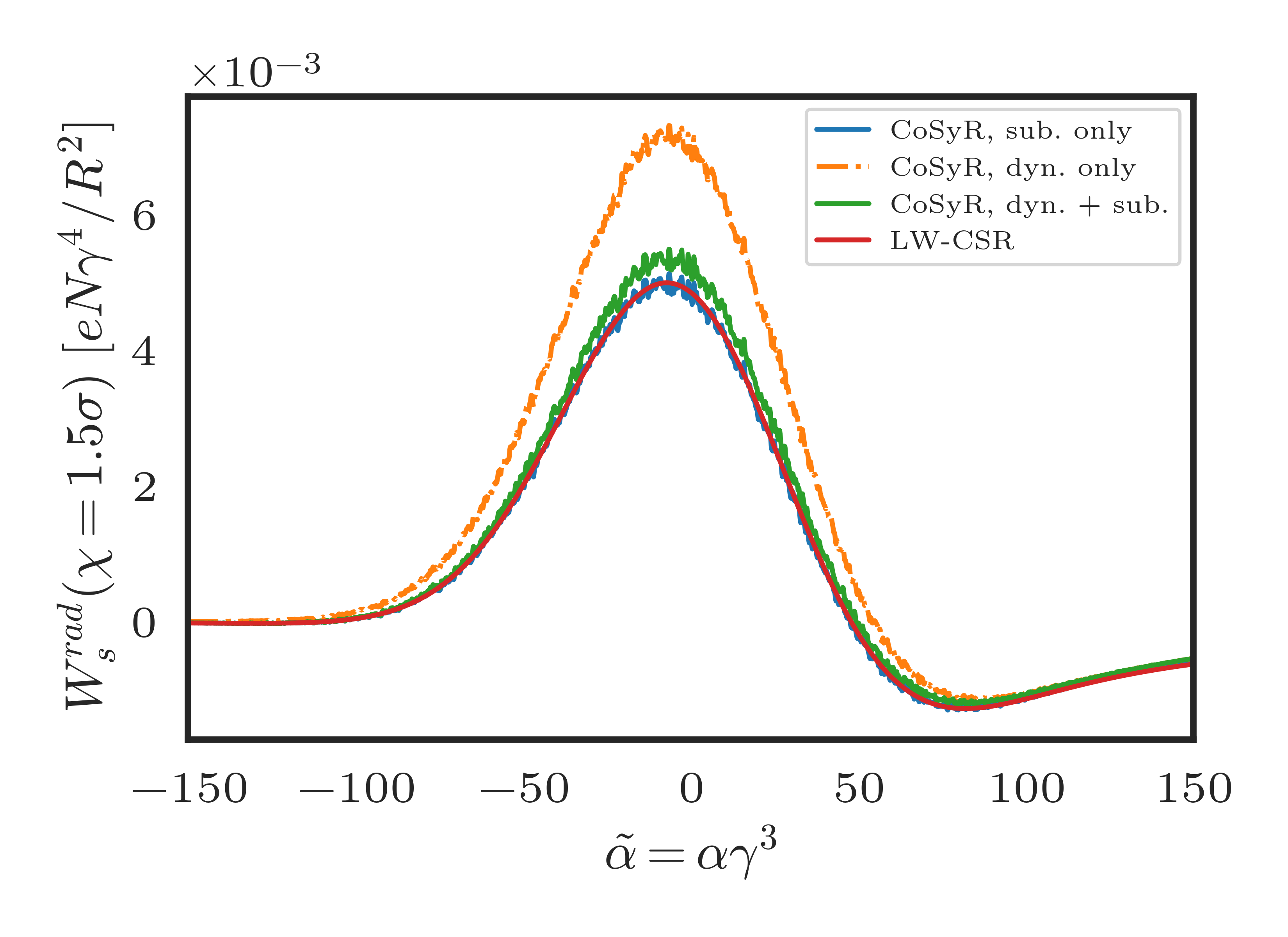} 
        \includegraphics[width=0.9\columnwidth]{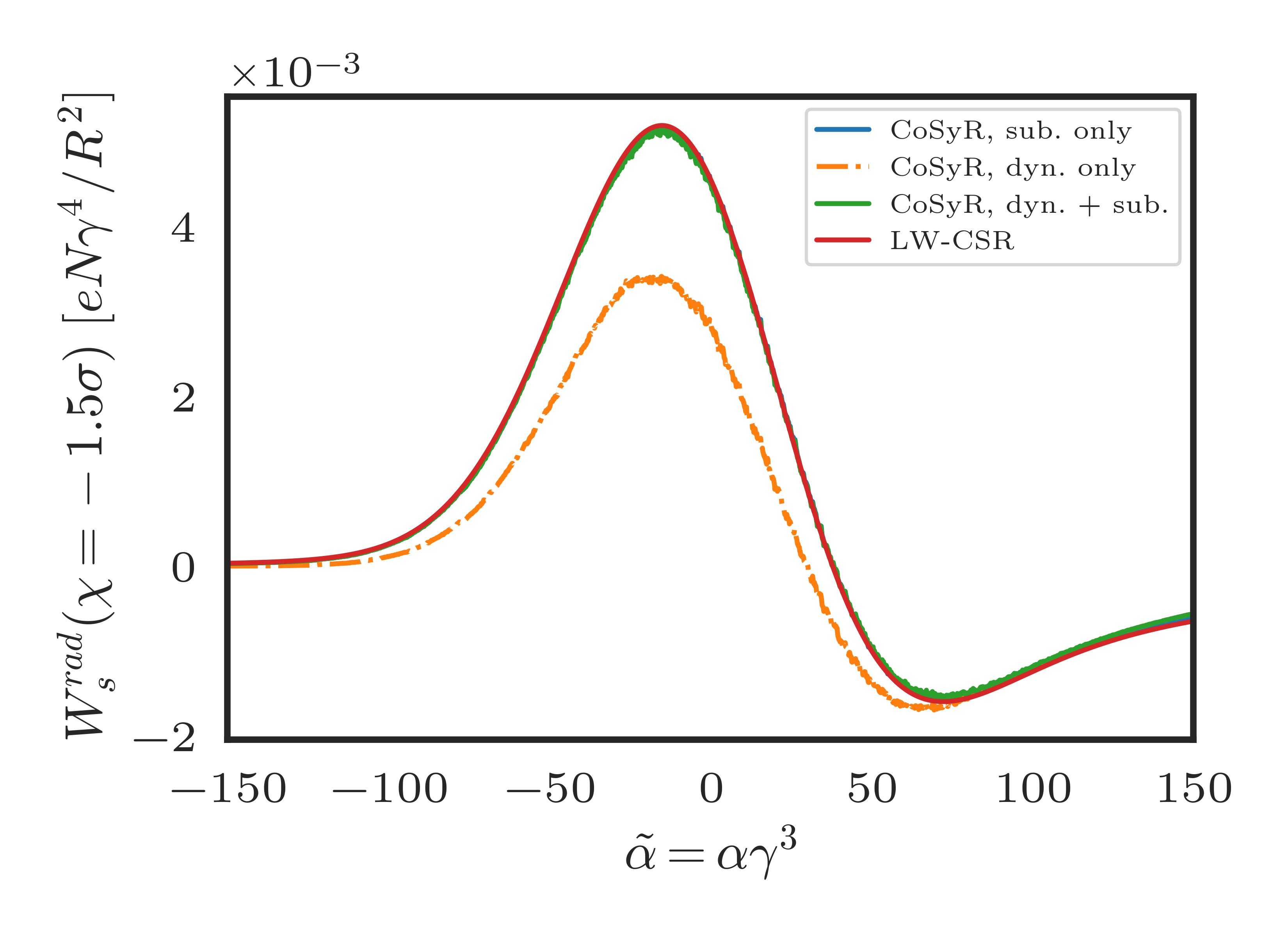} 
        \caption{The benchmark of the longitudinal coherent radiation field for the same setup as in Fig. \ref{fig:benchmark-g100}. The lineouts at $\chi=1.5\sigma$ (left) and $\chi=-1.5\sigma$  (right) are compared for three cases (1) with subcycle wavelets only, (2) with dynamic wavelets only and (3) with both dynamic and subcycle wavelets. The result obtained by convolving the field kernel with the beam density at a bending angle of $0.3$ rad is also shown (red solid curves denoted by "LW-CSR"), which is in good agreement with the subcycle wavelet only results. The difference in case (3) is most pronounced at the outer part ($\chi>0$) of the beam. No smoothing is done to the curves.}
        \label{fig:LW-CSR-benchmark}
\end{figure*}

As an example of the application of CoSyR, we demonstrate its capability through benchmarks of the longitudinal and transverse fields from the coherent synchrotron radiation and a beam dynamic simulation with such fields. Both 1D and 2D benchmarks have been carried out and here we focus on the 2D results which are more realistic than the 1D analytic analysis strictly valid for a Gaussian beam in steady-state.

In our benchmarks for the self-fields, only the external field is used to push particles. Both steady-state (using subcycle wavelets only to cover the entire moving mesh) and non-steady-state (using both dynamic and subcycle wavelets) results are shown and compared with those obtained from a 2D convolution procedure using the steady-state kernel and the instantaneous beam density \cite{huang2013two} (note that this is actually different from  the initial beam density due to bending).

Fig. \ref{fig:benchmark-g100} shows the steady-state longitudinal and transverse wakefields for an initial bi-Gaussian beam of $0.01$ nC charge, $\gamma=100$ and $33\mu$m spot size in a $R=1$m magnetic dipole bending section. Only subcycle wavelets are used to cover the entire simulation box and the snapshots are taken at a bending angle of $0.3$ rad which is sufficient to avoid the transient self-fields from the entrance. The boundary of the magnetic dipole region at the entrance has a sharp transition from the vacuum and refinements in time steps are used to resolve the beam motion in this transition region. The fields are normalized by $\gamma^4$ and the number of real electrons in the beam $N_p$. The transverse radiation electric field $E_{x}^{rad}$, which is the largest radiation electric field component, is also shown. However, the out-of-plane magnetic field from the radiation has a very similar profile and amplitude (not shown) leading to several orders of magnitude cancellation, and hence much weaker transverse wakefield $W_{\perp} = E_{x} - \beta_{0} B_{y}$.  The top half panels of the longitudinal and transverse wakefields are somewhat noisier than their lower half panels for the amount of the computation particles used, as a result of the spikes of the kernels located at the outer region from the trajectory.

Fig. \ref{fig:benchmark-g100-dyn} further shows the results when using both dynamic and subcycle wavelets in the simulation box. Subcycle wavelets are produced for the region with $\Psi<\Psi_{sub}=0.01$. This choice is made considering that the spike in the field kernel has a significant contribution to the coherent field on the beam scale, hence the spike should be covered only by the dynamic wavelets to allow a more realistic simulation as the beam evolves. For $\Psi_{sub}=0.01$ and the dynamic wavelet emission interval (which currently is an integral multiple of the pusher time step) used in the simulation, the subcycle and dynamic wavelets can overlap partially in the region away from the spike. By decreasing the threshold $\Psi_{sub}$ for the subcycle wavelets down to the value of $t_{sub}\beta$ where $t_{sub}$ is introduced in Eq. \eqref{t_sub}, the overlapping region will decrease but it is found that the result has converged for the threshold $\Psi_{sub}$ chosen.

Fig. \ref{fig:LW-CSR-benchmark} shows the detailed comparison of the longitudinal radiation electric field at two different transverse locations in the beam for three cases of wavelet coverage in CoSyR, namely, (1) with subcycle wavelets only, (2) with dynamic wavelets only and (3) with both dynamic and subcycle wavelets, as well as the steady-state convolution result (denoted "LW-CSR"). All results are qualitatively similar to the 1D analytic result \cite{MURPHY1997} and we have checked the agreement with the analytic result in 1D simulation setup for a Gaussian beam. Note that, as indicated by \cite{Saldin1997a,Derbenev1995}, the longitudinal coherent radiation field does work to the beam itself and it accelerates/decelerates the beam particles at the head/tail. For the results from case (1) and the convolution (i.e., “LW-CSR”), the field kernels are both approximated by ignoring the difference in individual particle's velocity. However, the former is obtained by remapping the kernel field from the subcycle wavelets and accumulating them on the mesh, while the latter is a direct convolution of the beam density with the field kernel. As can be seen, these two results are in good agreement, indicating that subcycle wavelets can represent the contribution from those emissions in the steady-state regime where the retarded time is smaller than the beam dynamic timescale. Nonetheless, the  CoSyR results have fluctuations corresponding to the number of computation beam particles ($\sim 10^7$) used in the simulation. Furthermore, variation of the longitudinal fields across the transverse dimension of the beam, i.e., deviation from the 1D scenario, is also observed. The case with dynamic wavelets only (orange curves) shows the largest variation across the beam transverse dimension (it also has a relatively large deviation with the 1D analytic result), while the results involving subcycle wavelets have overall smaller variation. This indicates that the dynamic wavelets inherently contribute to multi-dimensional effects missing in the 1D CSR models. For case (3) when both dynamic and subcycle wavelets are used (with small overlap outside the kernel spike region), the result represents contribution from all emissions both in the steady-state and non-steady-state regimes, hence more realistic than the other cases.

The effective transverse wakefield for the Gaussian beam has a non-uniform transverse profile as shown in Figs. \ref{fig:benchmark-g100} and \ref{fig:benchmark-g100-dyn}. But the difference between subcycle only and subcycle + dynamic cases appears to be small, likely due to the cancellation of the transverse electromagnetic fields in the transverse wakefield. It should be emphasized that the overall field still contains contributions from steady-state and non-steady-state emissions which are associated with two different timescales.

\begin{figure*}[ht!]
        \centering
        \includegraphics[width=0.68\columnwidth]{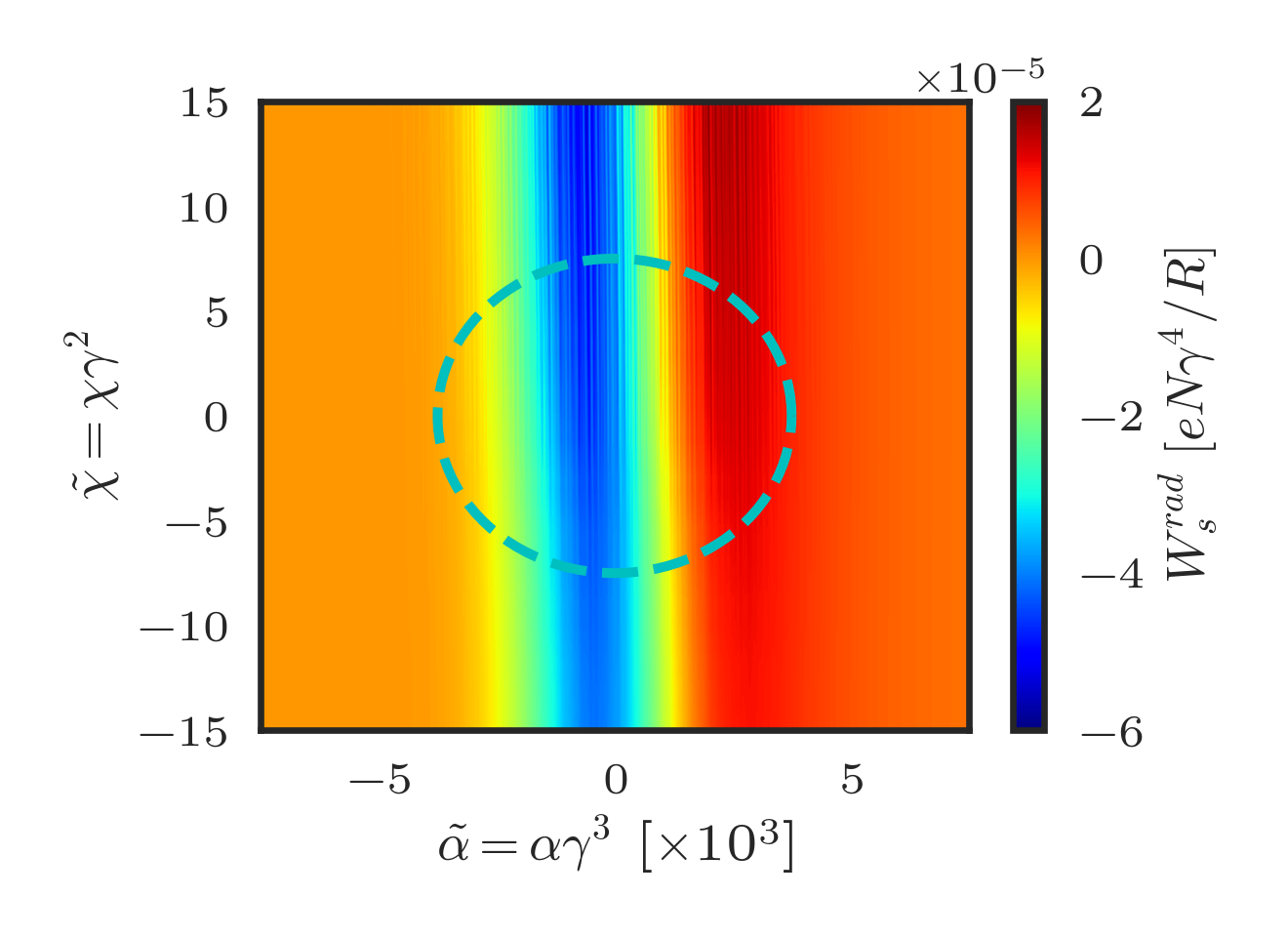}
        \includegraphics[width=0.68\columnwidth]{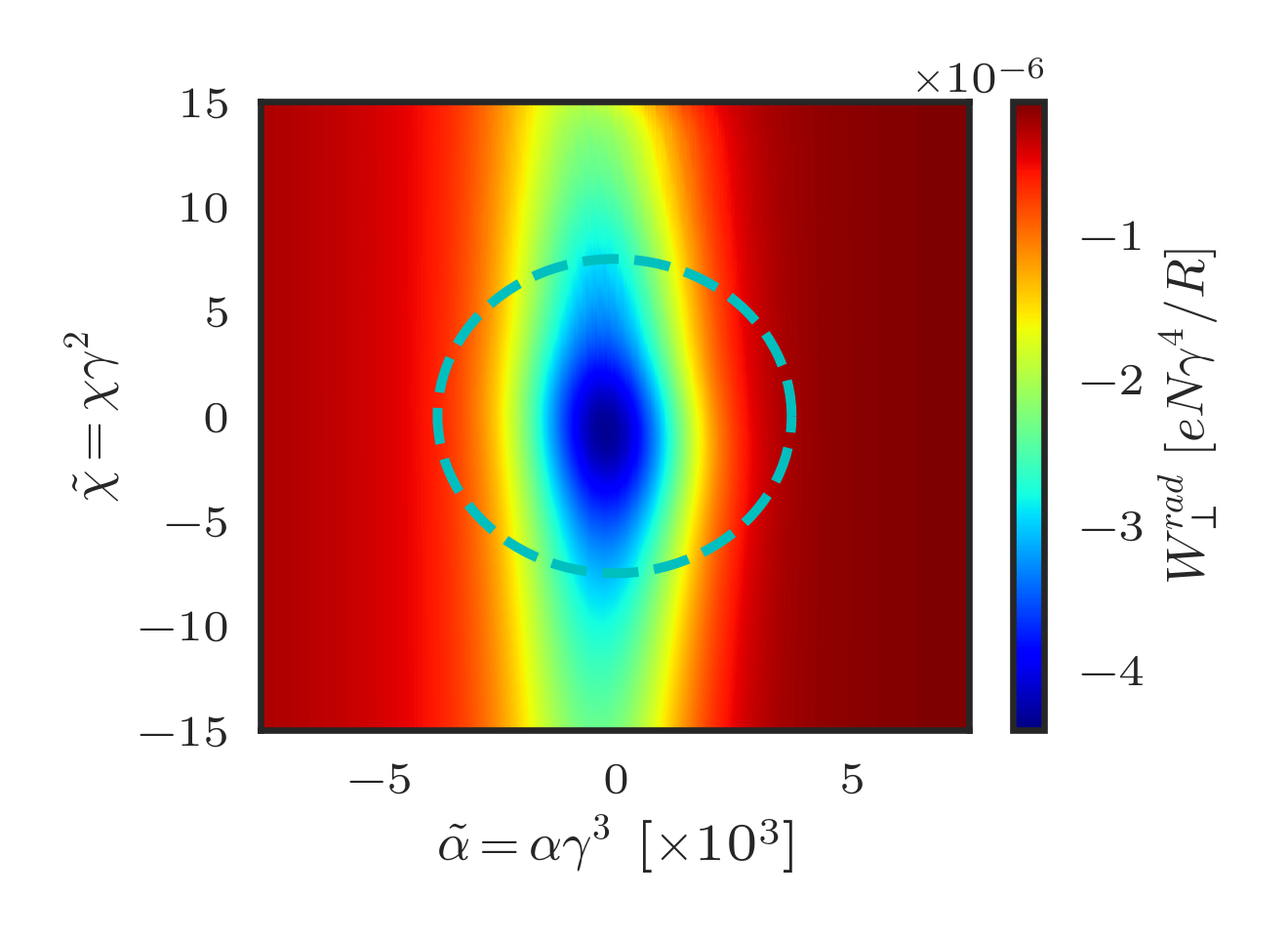}
        \includegraphics[width=0.68\columnwidth]{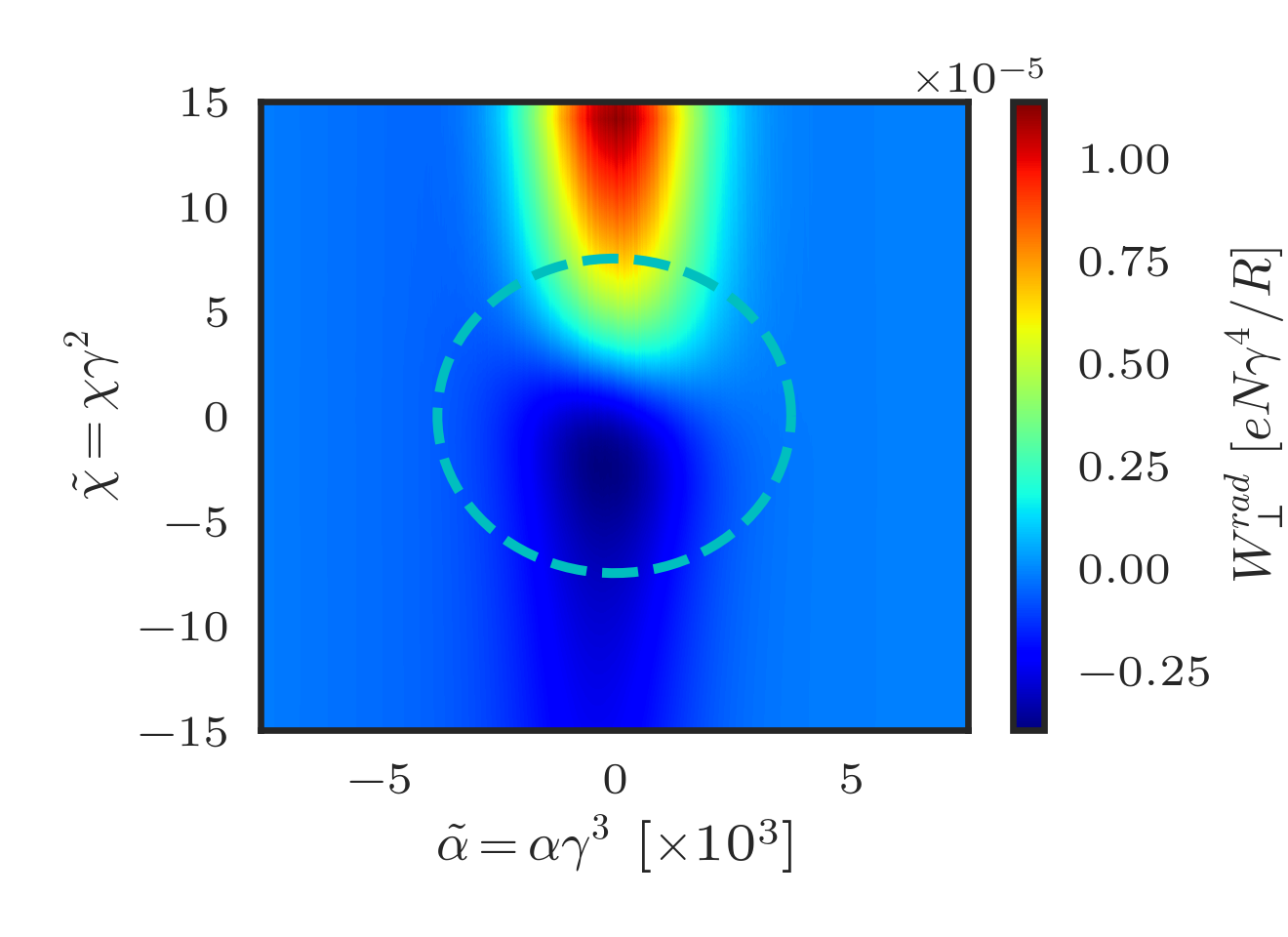}        
        \caption{The longitudinal wakefield (left), the effective transverse wakefield (middle, for the case with subcycle only; right, for the case with dynamic + subcycle wavelets) from the coherent radiation (acceleration field) for a beam with initial $\gamma=500$ and spot size of $10 \mu \text{m} \times 10 \mu \text{m}$ in a magnetic dipole bend. The bending radius is $R=1$m and the beam is at $0.2$ rad into the bend. Both dynamic and subcycle wavelets are used with the “mixed-kernel” formalism. The longitudinal wakefield in the bending plane (bi-linear smoothing is used for the contour plot) is obtained by a least squares fit estimate of the gradient of the pseudo-potential $(\phi - \beta A_{s})$. $\sim 2\times 10^6$ computation particles are used for the beam.}
        \label{fig:benchmark-g500-dyn-sub-long-trans-wake}
\end{figure*}

\begin{figure*}[ht!]
        \centering
        \includegraphics[width=0.9\columnwidth]{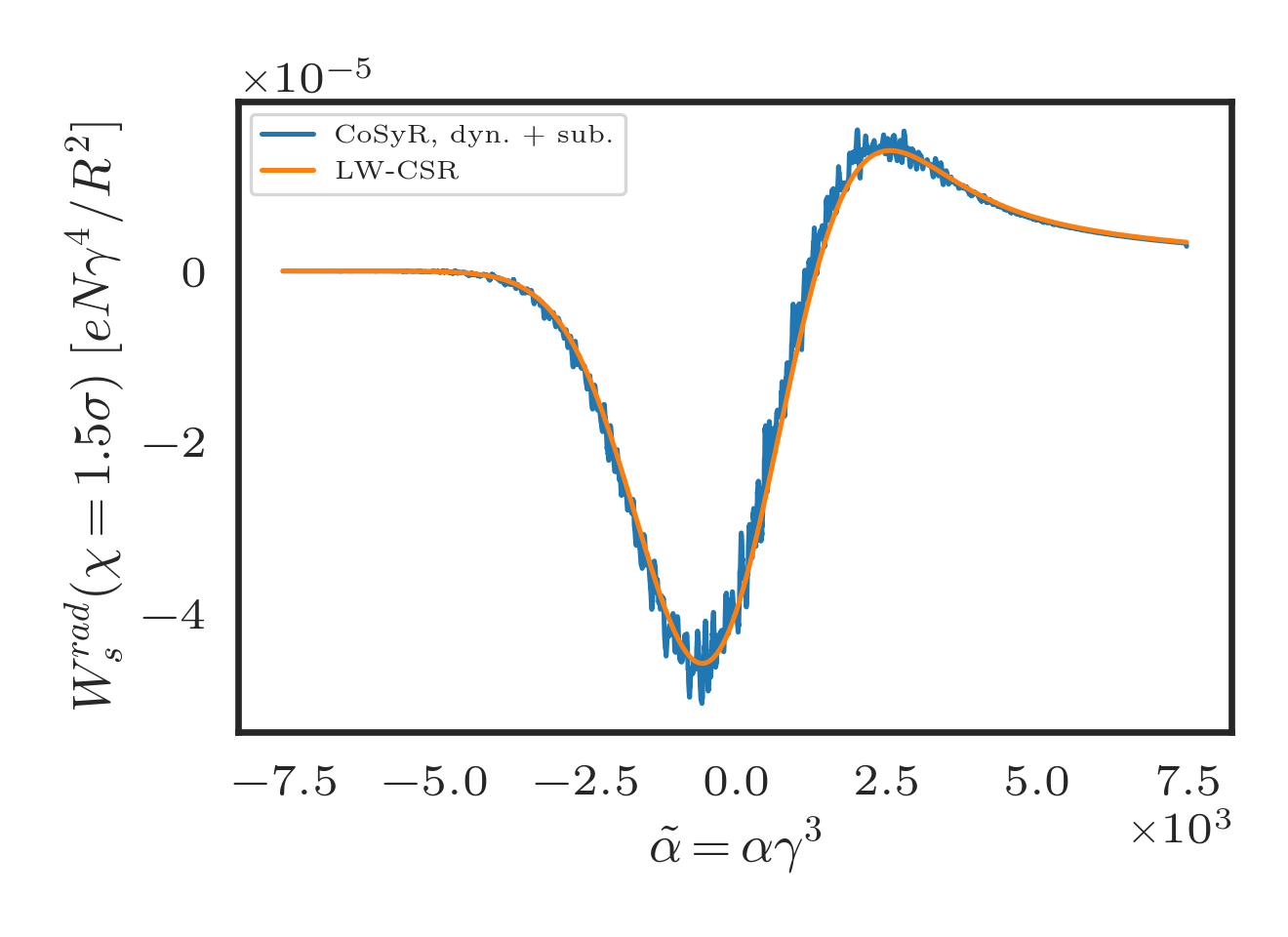}
        \includegraphics[width=0.9\columnwidth]{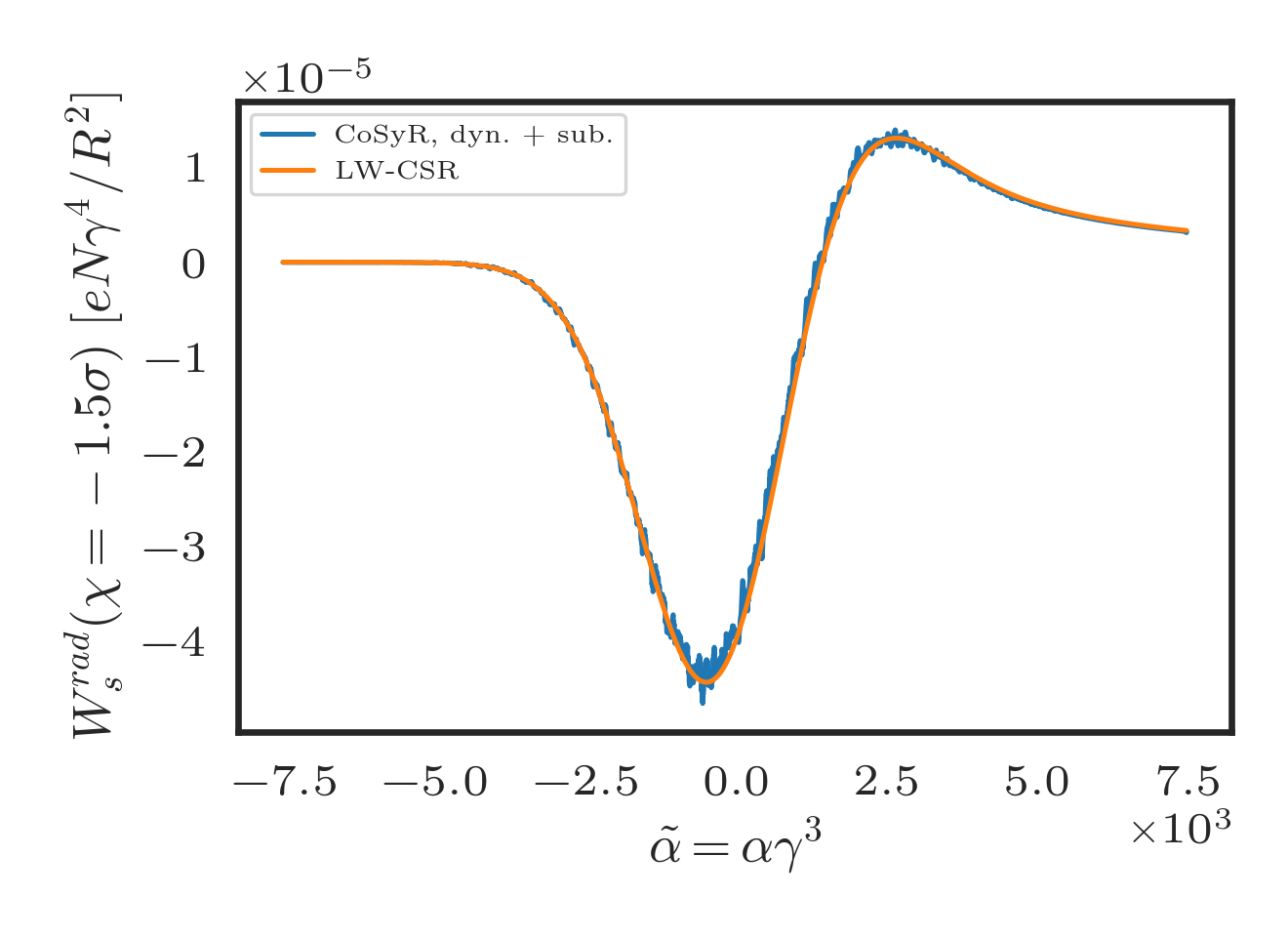}   
        \caption{Comparison of the longitudinal wakefield field lineouts at two transverse locations $\chi=\pm1.5 \sigma$ with the convolution result (“LW-CSR”).  No extra smoothing is used for the lineouts.}
        \label{fig:benchmark-g500-dyn-sub-long-wake}
\end{figure*}

To demonstrate the “mixed-kernel” formalism, a benchmark with a higher energy beam is conducted. In this benchmark, the beam Lorentz factor is $\gamma=500$ and the beam spot size is $10 \mu \text{m} \times 10 \mu \text{m}$. The bending radius is kept the same ($R=1$m). Fig. \ref{fig:benchmark-g500-dyn-sub-long-trans-wake} shows the longitudinal wakefield, and Fig. \ref{fig:benchmark-g500-dyn-sub-long-wake} shows the comparison between the case with dynamic + subcycle wavelets and the result from convolution for the longitudinal wakefield. Subcycle wavelets are produced for the region where $\Psi<\Psi_{sub}=0.002$. The agreement between the two is good except for the fluctuations in the CoSyR result due to the relatively small amount ($\sim 2\times 10^6$) of computation particles used. At this beam energy, the variation of the longitudinal wakefield across the transverse dimension of the beam is small, therefore the 1D analytic result can be a good approximation. We have checked that the longitudinal wakefield result using only subcycle wavelets (not shown) also has similar agreement as expected.

Fig. \ref{fig:benchmark-g500-dyn-sub-long-trans-wake} also shows the comparison of the effective transverse wakefield between the case with subcycle only and the case with dynamic + subcycle wavelets. For the former case, the result is qualitatively similar to the $\gamma=100$ beam benchmark with the “field-only” formalism. Furthermore, we have verified that the result from the convolution method is also in agreement for this effective transverse wakefield. However, the result with dynamic + subcycle wavelets is substantially different from the results employing the steady-state assumption. The shape of the effective transverse wakefield resembles that of the transverse components of the electromagnetic fields (see $E_{x}^{rad}$ in Fig. \ref{fig:benchmark-g100}), which is likely a result of an incomplete cancellation of electromagnetic field components in the effective transverse wakefield due to the deviation from the reference trajectory.

\subsection{Beam dynamics}

\begin{figure*}[htbp]
        \centering
        \includegraphics[width=0.9\textwidth]{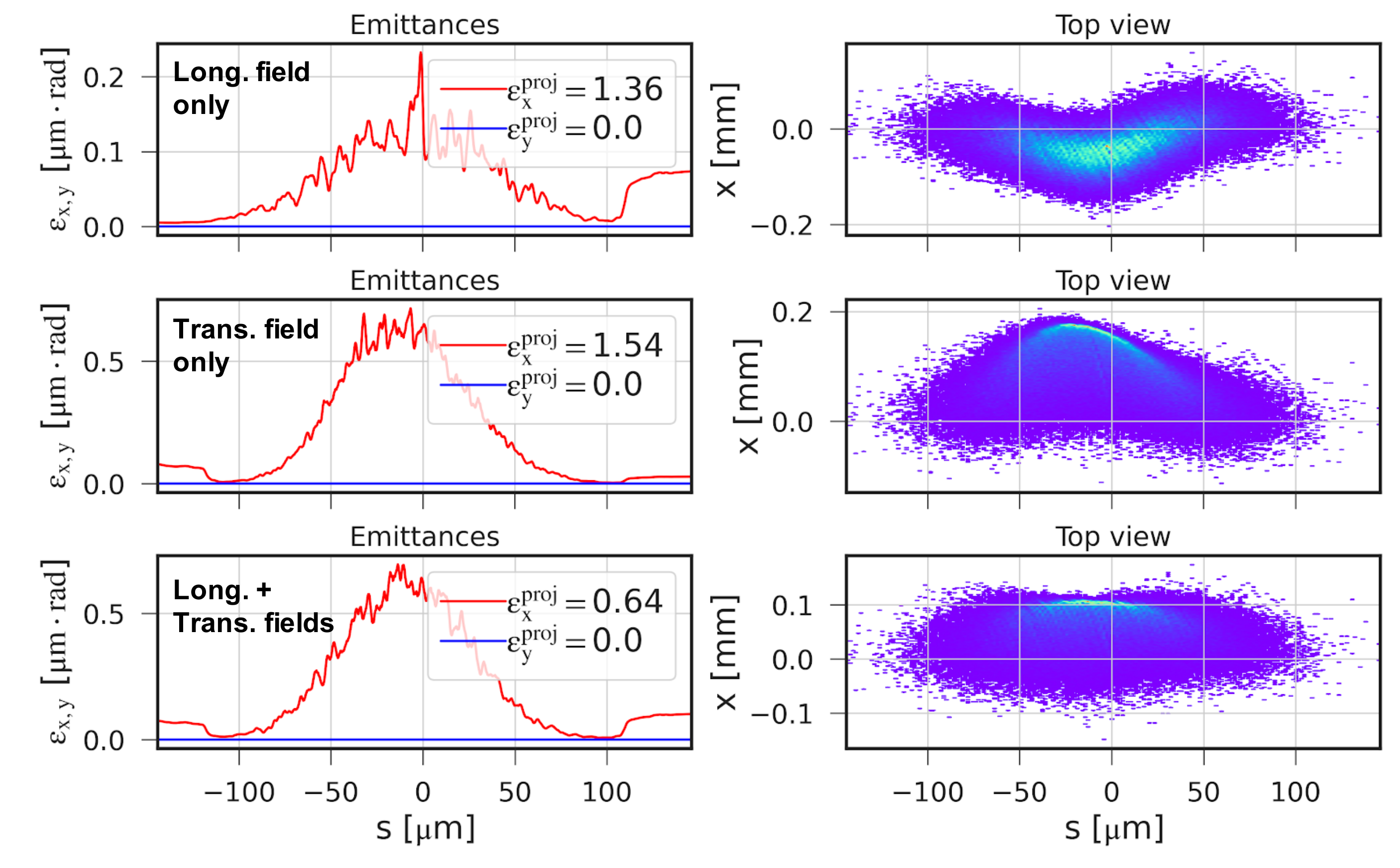} 
        \caption{The slice emittance (solid lines in left column) and the shape of an electron beam (right column) after propagation for $0.19$rad in a circular trajectory of radius $R=1$m. The initial beam is round and has emittance $\epsilon_{x,y}=0$ and a spot size of $30\mu$m with peak current $\sim 3k$A. Three 2D simulations are compared to show the roles of the longitudinal and transverse (radial direction denoted by $x$) radiation field and the effect of their interplay: (top row) result with longitudinal field only, (middle row) result with transverse field only, (bottom row) result with both fields.  The projected beam emittances are indicated by  legends in each case.}
        \label{fig:csr_beam_dynamics}
\end{figure*}

Further beam dynamic simulations using CoSyR are shown in Fig. \ref{fig:csr_beam_dynamics}. This is conducted self-consistently without the usual steady-state assumption for the first time as far as we are aware of. More detailed simulation study of the beam dynamics evolution will be presented in a separate publication. Here we briefly discuss the simulation results that reveal and confirm a complex interplay between the longitudinal and transverse coherent fields by the beam itself. In the simulations, an initially round beam with bi-Gaussian profile of $33\mu$m spot size and $\sim 3$kA peak current is simulated in a $R=1$m magnetic dipole bend. The initial Lorentz factor is $\gamma = 100$.  While the energy gain/loss from the longitudinal field leads to an $s$-shape distortion for the beam circulating in the magnetic dispersive section (upper right panel), the transverse field introduces opposite motion to offset the distortion (middle right panel). This results in a residual net displacement of the beam centroid (Fig. \ref{fig:csr_beam_dynamics} bottom right panel). Interestingly, this also leads to smaller projected emittance growth of the beam than the cases with the longitudinal or transverse field alone. However, the slice emittance growths are comparable for the last two simulations indicating that it is mostly due to the transverse field.

\section{Conclusion}

In this work, we have developed a novel beam dynamics simulation code, CoSyR, for the modeling of self-consistent synchrotron radiation effects on a high brightness beam. Modeling near-field synchrotron radiation accurately and efficiently is a critical challenge to the understanding of beam dynamics in many advanced light sources and accelerator concepts. Although many methods and simulation tools, which are mostly based on 1D or multi-dimensional steady-state models of the coherent synchrotron radiation in a circular trajectory, have been developed in the past, the pursuit of a versatile and accurate modeling tool is still an active area. In particular, self-consistent multi-dimensional simulations are lacking, which are important to assess the beam instabilities and emittance growth in future facilities at the frontier of accelerator physics.

Our code is based on the Green's function of Maxwell's equations, i.e., the Liénard-Wiechert formalism, and the concept of wavefront propagation first proposed by Shintake. We have designed an algorithm to dynamically choose the wavelets on these wavefronts for the evaluation of the kernel fields/potential. Two types of kernel choices are experimented and implemented for relatively low energy beams, while further direction towards a kernel with better convergence property, e.g., as in Ref. \cite{Cai2020} is worth investigating. Such an option can be implemented in our code in a straightforward manner and will significantly enhance the accuracy and efficiency for modeling high energy beams. For close-by emissions, subcycle wavelets are used and the kernel evaluation is simplified assuming the dependence only on the reference particle velocity. The different beam dynamic timescales are essentially separately handled by the dynamic and subcycle wavelets.

We further utilize a parallel remapping library, Portage, to construct the remapped field from the wavelets onto a moving mesh. The other modules, including kernel calculation and particle update, are implemented using Kokkos and Cabana to allow easy portability to future Exascale platforms.

The benchmark for both longitudinal and transverse wakefields and the beam dynamics in the coherent fields are demonstrated, where some detailed understanding are briefly discussed. This capability opens up the opportunity for first-principle study of the interaction of high brightness beams with their radiation fields. In the future, this will be improved with better kernel options, more flexible simulation setup, inclusion of incoherent radiation effects and the integration with global simulations with external boundaries and cavities.

\section{Acknowledgements}

Research presented in this article was supported by the Laboratory Directed Research and Development program of Los Alamos National Laboratory under project number 20190131ER. This research used resources provided by the Los Alamos National Laboratory Institutional Computing Program, which is supported by the U.S. Department of Energy National Nuclear Security Administration under Contract No. 89233218CNA000001. In addition, this research used resources of the National Energy Research Scientific Computing Center (NERSC), a U.S. Department of Energy Office of Science User Facility located at Lawrence Berkeley National Laboratory, operated under Contract No. DE-AC02-05CH11231.

\appendix*
\section{}

\begin{figure*}[ht!]
        \centering
        \includegraphics[width=0.9\textwidth]{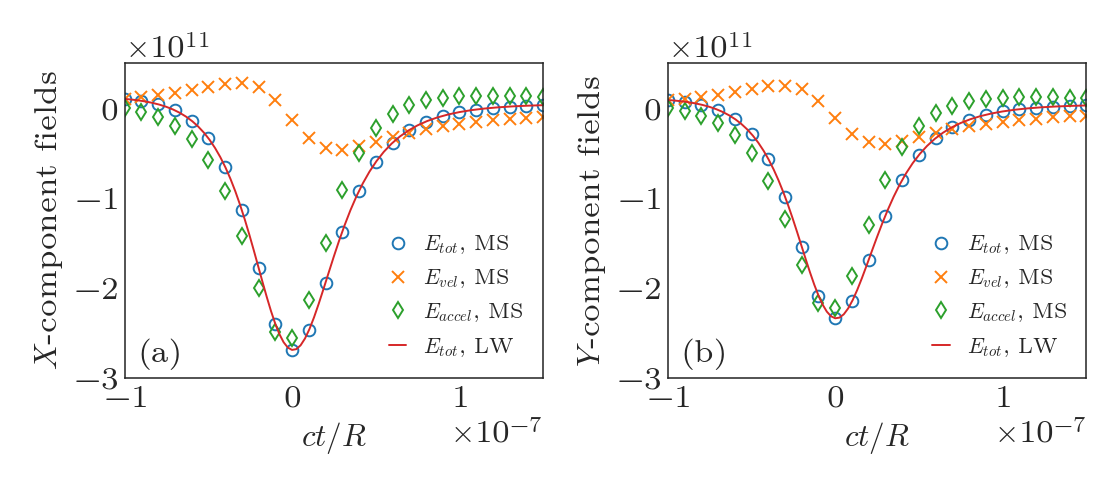}
        \caption{Comparison of the fields calculated from the modified Shintake (MS) scheme and the Liénard-Wiechert (LW) equation. The electron of $\gamma=200$ starts from the top of a circular trajectory of radius $R$. The observation point is radially off the trajectory by $10^{-4}R$ and at an angle of 50 degrees relative to the vertical axis. (a,b) compare the $x, y$ field components, respectively. The fields are presented in arbitrary unit.
        }
        \label{near-field-compare}
\end{figure*}

While the original idea of Shintake was mainly used to animate the fieldline pattern in the near-field zone, it also pointed out a seemingly convenient way to calculate the fields at the corresponding wavelets. The central argument was based on applying the Gauss law to a moving charge. It suggested that the field flux enclosed by a given pipe should be determined by the portion of electric charge that contributed to the flux and the flux should be a constant regardless of the ensuing electron motion. Therefore, this argument implies a simple method for field calculation by measuring the wavelet density and their distance to the origin of emission; both are tracked in the simulation and can be easily obtained. However, a closer investigation ~\cite{li2019validation} shows that the above argument is valid only for a linear motion with no acceleration. Although a general motion can be numerically discretized into small segments of uniform linear motion, the field calculation still has to take into account both the velocity field and the acceleration field even in the instantaneous rest frame of the electron (denoted by double primes), i.e.,

\begin{equation}
        \label{vel_acc_fields}
        \vec{E}_{vel}''=-e\frac{\hat{n}''}{r''^2}, \vec{E}_{accel}''= - e\frac{\hat{n}''\times(\hat{n}''\times \vec{a}'')}{r''}.
\end{equation}

Here, $\hat{n}''$ is the unit vector pointing from the emitting point to the field point, $r''$ is the distance between the two points, and $\vec{a}''$ is the acceleration. These fields in the instantaneous rest frame of the electron can then be Lorentz transformed back to the laboratory frame. The transform is found~\cite{Singal2011} to reproduce exactly the Liénard-Wiechert equation in the lab frame (retarded distance and velocity denoted by prime),

\begin{equation}
        \label{lw-eq}
                \vec{E} = \frac{-e(\hat{n} - \vec{\beta}')}{\gamma^2 r'^2 (1-\hat{n}\cdot\vec{\beta}')^3}+\frac{-e\hat{n} \times [(\hat{n}-\vec{\beta}') \times \dot{\vec{\beta}}'] }{ c r'(1-\hat{n}\cdot\vec{\beta}')^3}.
\end{equation}

We have numerically checked the double Lorentz transform~\cite{li2019validation} by calculating the fields at a fixed point near the electron cyclotron path, and the resulting velocity and acceleration fields are displayed separately in Fig.~\ref{near-field-compare} for both the $x, y$ components. Detailed parameters of the setup can be found in the caption. It is seen that for the cyclotron motion, the acceleration field can dominate over the velocity field, and their sum reproduces the calculation using the Liénard-Wiechert equation directly. The double Lorentz transform involves a considerable amount of computation; therefore, CoSyR uses the Liénard-Wiechert equation to save computation.